\documentclass{amsart}
\usepackage{amsmath,amssymb,amsthm,amsfonts}
\usepackage[utf8]{inputenc}
\usepackage{amsmath}
\usepackage{graphicx}
\usepackage{xcolor}
\usepackage{marginnote}
\newcommand{\RR}{\mathbb R}
\newcommand{\ZZ}{\mathbb Z}
\newcommand{\pa} {\partial}

\begin{document}
\title{The two-point correlation function \\ in the six-vertex model}

\author{Pavel Belov}
\address{P.B.: Spin Optics Laboratory, St. Petersburg State University, Ulyanovskaya 1, St. Petersburg, 198504 Russia}
\email{pavelbelov@gmail.com}

\author{Nicolai Reshetikhin}
\address{N.R.: Department of Mathematics, University of California, Berkeley,
CA 94720, USA \& Department of Physics, St. Petersburg State University, Ulyanovskaya 1, St. Petersburg, 198504 Russia \& KdV Institute for Mathematics, University of Amsterdam, Science Park 904, 1098 XH Amsterdam, The Netherlands.}
\email{reshetik@math.berkeley.edu}

\maketitle

\begin{abstract}
We study numerically the two-point correlation functions of height functions in the six-vertex model with domain wall boundary conditions. The correlation functions and the height functions are computed by the Markov chain Monte-Carlo algorithm. Particular attention is paid to the free fermionic point ($\Delta=0$), for which the correlation functions are obtained analytically in the thermodynamic limit. A good agreement of the exact and numerical results for the free fermionic point allows us to extend calculations to the disordered ($|\Delta|<1$) phase and to monitor the logarithm-like behavior of correlation functions there. For the antiferroelectric ($\Delta<-1$) phase, the exponential decrease of correlation functions is observed.
\end{abstract}

\section{Introduction}

\begin{figure}[ht!]
\begin{center}
\includegraphics[width=1.0\textwidth, angle=0.0]{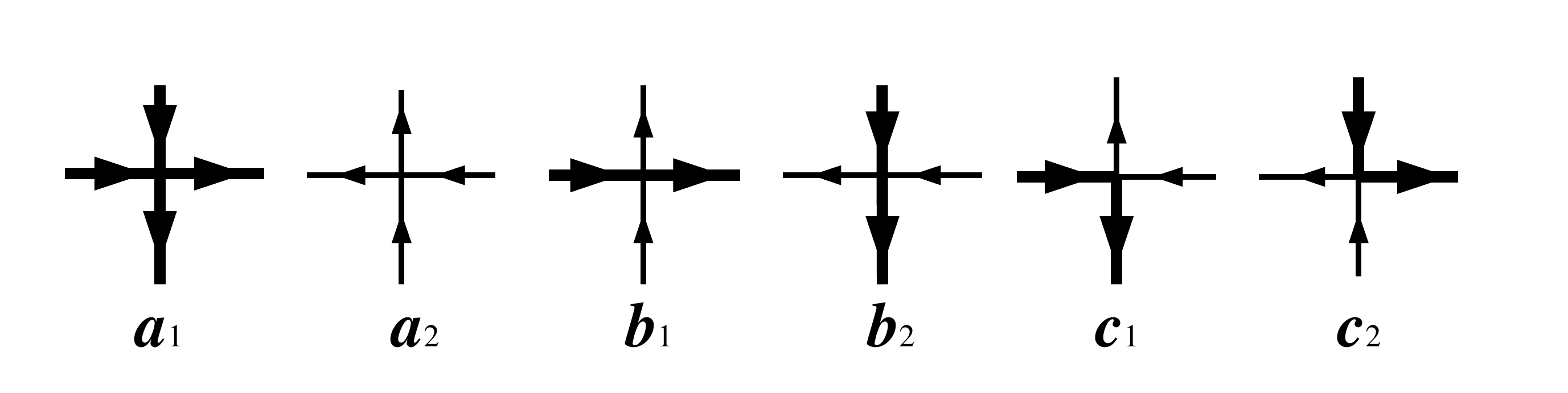}
\end{center}
\caption{\label{6v} Local configurations and weights of the six-vertex model. For the symmetric model, $\tau=\tau_{1}=\tau_{2}$, where $\tau=a,b,c$.}
\end{figure}

The goal of this paper is the computation of correlation functions in the six-vertex model directly by Markov chain Monte-Carlo simulations. This model was introduced by Pauling~\cite{Pauling} who proposed it to describe the crystal where the oxygen groups form a square lattice with a hydrogen atom between each pair of lattice sites.
He proposed the ice rule: each lattice site has two hydrogen atoms close to it and two further apart; see a recent historical review~\cite{Harris}.
Another crystal with such a structure is the potassium dihydrogen phosphate.
Slater was the first who suggested that the two dimensional case, known as the six-vertex model, is important 
to understand universal thermodynamic properties of these structures~\cite{Slater}. 
The states in this model are configurations of arrows on edges which satisfy the ice rules, see Fig.~\ref{6v}.
An arrow indicates to which of two sites (the oxygen atoms) the hydrogen atom (which is approximately in the middle of an edge) is closer.
Equivalently, the configurations of arrows can be regarded as configurations of lattice paths such that paths may meet at a vertex, 
turn or pass, as it is shown in Fig.~\ref{6v}. Boltzmann weight of a configuration is the product of Boltzmann weights assigned to vertices. The weight of a vertex depends on the configurations of paths on adjacent edges,
see Fig. \ref{6v}. The probability of state $\sigma$ is
\[
Prob(\sigma)=\frac{1}{Z} \, w(\sigma),
\]
where $w(\sigma)=\prod_v w_v(\sigma)$ is the weight of  state $\sigma$, $w_v(\sigma)$ is the weight of the vertex $v$
in the state $\sigma$, and $Z=\sum_\sigma w(\sigma)$ is the partition function. 

Locally, lattice paths of the six-vertex model on a planar lattice can be regarded as level curves of a step function
defined on faces~\cite{R2010}. We assume it is increasing when we move to the right and up. This integer valued function is called 
the {\it height function} $\chi(n,m)$. It is a random variable with values in integers $\ZZ$ with the probability distribution given by Boltzmann weights described above.
On a planar simply connected lattice domain there is a bijection between configurations of paths with fixed positions on the boundary and height functions with corresponding boundary values\footnote{We assume that the value of a height function is fixed at some reference point on the domain.}. 

The first breakthrough in the study of the model came in works of Lieb, Yang, Sutherland and others
where the Bethe ansatz was used for finding the spectrum of transfer matrices with periodic boundary conditions~\cite{Lieb,Yang,Sutherland}.
Then, came works of Baxter where the role of commuting transfer matrices became clear and the partition function of the eight-vertex model was obtained~\cite{BaxterPaper}, see~\cite{Baxter} for an overview of these developments.
Then, many important algebraic structures came in the framework of the algebraic Bethe ansatz and the quantum inverse scattering method~\cite{FT}, for an overview see~\cite{BIK,BM,AGP}.
In the last decade, substantially better understanding of thermodynamic properties of the six-vertex model with domain wall boundary 
conditions (DW) on a square lattice has been achieved.
These boundary conditions correspond to paths coming through each edge on the top side of the square and leaving through edges on the right side only.
These particular boundary conditions are quite remarkable because the partition function in this case can be written as a determinant~\cite{Izergin,BIK,PronkoTMF} as well as because of the relation to 
the alternating sign matrices~\cite{Ku}.

In the large volume limit (the thermodynamic limit, $N\to \infty$), properly normalized height function converges, as a random variable, 
to a deterministic function $h_{0}(x,y): D=[0,1]\times [0,1]\to \RR$ known as the limit shape
height function.
Such a behavior is known as the limit shape phenomenon, see~\cite{CP,Ok} for an overview.
This phenomenon is an analogue of the central limit theorem in probability theory.
It predicts the following behavior of the height function as $N\to \infty$\footnote{This means the convergence of $\chi(n,m)/N\to h_0\left(\frac{n}{N}, \frac{m}{N}\right)$ and $\chi(n,m)-Nh_0\left(\frac{n}{N}, \frac{m}{N}\right)\to \phi\left(\frac{n}{N}, \frac{m}{N}\right)$ in probability.}:
\begin{equation}
\label{eq1chi}
\chi \left(n,m \right) \to N h_0\left( \frac{n}{N}, \frac{m}{N} \right)+\phi \left(\frac{n}{N}, \frac{m}{N} \right).
\end{equation}
Here $h_0(x,y)$  is the limit shape height function which can be computed using the variational principle.
The variational principle was developed and proved for dimer models in~\cite{KOS,Ok}. It was adopted to the six-vertex model in~\cite{ZJ,PR}.
The random variable $\phi(x,y)$ is a free Gaussian quantum field in the Euclidean space time with the metric determined by the height function $h(x,y)$, see for example~\cite{ZJ,GorinLectures}.
For generic values of parameters in the six-vertex model, mathematically, the variational principle is still a hypothesis~\cite{PR,GBDJ}. It is proven in some special cases of stochastic weights in~\cite{B-stoch} and
it follows from~\cite{CKP} for the free fermionic case $\Delta=0$. 

An important characteristic of the (symmetric) six-vertex model is the parameter~\cite{BaxterPaper}
\begin{equation}
\label{delta}
\Delta=\frac{a^2+b^2-c^2}{2ab}.
\end{equation}
When $\Delta=0$, the model can be mapped to a dimer model and the partition function and correlation functions
can be computed in terms of the determinant and the minors of the Kasteleyn matrix~\cite{Ka}, respectively. 

It has already been shown earlier~\cite{R2005,AS} how to use configurations generated by a Markov chain Monte-Carlo simulations for calculating the limit shape of the height function of the six-vertex model with DW boundary conditions.
In this paper, we show how, based on the generated configurations, to compute numerically the two-point correlation function.
When $\Delta=0$ both the limit shape height function and the correlation functions are known from the exact solution because in this case the six-vertex can be mapped to a dimer model on a modified (decorated) square lattice, details can be found in~\cite{PR,RS}. We demonstrate that in this case the usual averaging over time in Markov process gives an excellent agreement of numerical results with the exact ones.

After that, we apply the same algorithm for other values of $\Delta$.
When $|\Delta|\leq 1$ the model is critical, i.e. the Gaussian field $\phi(x,y)$ is a massless 
field on the space time with the metric induced by the limit shape. The numerics confirms that 
correlation functions are conformal at short distances. Of course, we should not expect conformal invariance at all distances
for $\Delta$ other than zero. 

When $\Delta<-1$ an antiferroelectric diamond shape droplet forms
in the middle of the limit shape. Because the antiferroelectric ground state is double degenerate,
the Markov process gets stuck in one of the ground states for exponentially long time. 
Numerical estimations for this case are given in the last section.

The paper is organized as follows. In section~\ref{exact}, we outline a derivation of the
two-point correlation function from the exact solution of the six-vertex model for $\Delta=0$.
In section~\ref{num}, we demonstrate the results of numerical Monte-Carlo simulations and comparisons with the exact solution.
In the appendices, we provide the technical details to specify the model to obtain the exact solution.
 
{\bf Acknowledgements.} We would like to thank A.~G.~Pronko for discussions and for sharing a draft of the manuscript~\cite{Pronko},
D. Keating and A. Sridhar for numerous discussions and the latest version of the Monte-Carlo code.
We also benefited from discussions with A.~A.~Nazarov.
The work of N.~Yu.~Reshetikhin was partly supported by the
NSF grant DMS-1902226 and the RSF grant 18-11-00297.
P.~A.~Belov is grateful to the Russian Science Foundation, grant no. 18-11-00297, for the financial support.
The calculations were carried out using the facilities of the SPbU Resource Center ``Computational Center of SPbU''.

\section{Correlation functions in the six-vertex model at the free fermionic point}\label{exact}

\subsection{The free fermionic point of the six-vertex model and mapping to dimers}\label{LSh}
In this paper, we focus on the symmetric six-vertex model with weights $a_{1}=a_{2}=a$, $b_{1}=b_{2}=b$, $c_{1}=c_{2}=c$.
The parameter $\Delta$, Eq.~(\ref{delta}), is an important characteristic of the model.
It determines the phases of the model on the $M\times N$ torus when $M,\, N \to \infty$.
For simplicity, in the following we assume that $M=N$.
When $\Delta>1$ the model develops a ferroelectric, totally ordered phase.
For $|\Delta|<1$ it develops a disordered phase and for $\Delta<-1$ it transitions to an antiferroelectric phase.
When $\Delta=\pm 1$ the model undergoes phase transitions (in parameter $\Delta$).

When the weights of the six-vertex model satisfy the condition $\Delta=0$
the six-vertex model can be mapped to the dimer model on a modified lattice, see for example
\cite{Wu,RS}.
The partition function of the dimer model is the sum of Pfaffians (the number of terms is determined by the topology of the lattice)~\cite{Ka,FisherPR,McCoyWu}.
Each Pfaffian can be regarded as the Gaussian Grassmann integral.
Because of this and because it implies that the multipoint correlators can be expressed as Pfaffians of the two-point correlation functions, the case of $\Delta=0$ is also known as a free fermionic point of the six-vertex model.

Because the weights can be multiplied by an overall constant factor without changing the probability measure, we can put $c=1$.
Then, we can parametrize weights $a$ and $b$ as
$$
a=\cos{(u)}, \enskip b=\sin{(u)}.
$$
This is a particular case of Baxter's parametrization of 
weights of the six-vertex model \cite{Baxter}. 
Note that the mapping $a\mapsto b, \ \ b\mapsto a$ is a symmetry of the 
probability measure, see Appendix \ref{Sym} for details. This is why we can assume,
without loosing generality, that $b/a\leq 1$. 

\subsection{The variational principle}
Here we will recall the variational principle for deriving the limit shape.
Let $\sigma(s,t)$ be the free energy per site for the six-vertex model on a torus with $s$ and $t$ being fixed
densities of edges occupied vertical and horizontal paths respectively.

The limit shape height function $h_0(x,y)$ for the six-vertex model with DW boundary conditions is a real valued 
function on $\mathcal D=[0,1]\times [0,1]$ which minimizes the large deviation rate functional
\begin{equation}\label{LDf}
S[h]=-{\int \int}_{\mathcal D} \sigma(\pa_xh,\pa_yh) \, dx \, dy
\end{equation}
in the space of functions with boundary conditions $h(0,y)=h(x,0)=0, \ \ h(1,y)=y, \ \ h(x,1)=x$ and the constraints
\[
|h(x,y)-h(x',y)|\leq |x-x'|,  \ \ |h(x,y)-h(x,y')|\leq |y-y'|.
\]
For dimer models it follows from~\cite{CKP}.

The critical value $S[h_{0}]$ is the minus free energy of the model.
If $Z_{N}$ is the partition function of the six-vertex model with DW boundary conditions, then
$$
S[h_{0}] = \lim_{N\to\infty} \frac{1}{N^{2}} \ln Z_{N}.
$$

The six-vertex model at the free fermionic point~($\Delta=0$) can be mapped to the dimer model 
on a decorated square lattice, see for example~\cite{RS} and references therein.
As it was already stated earlier, the corresponding dimer model can be solved by the Pfaffian method.
This method gives the formula for the partition function of the model as a Pfaffian (or a determinant) of certain $N\times N$
matrix, called the Kasteleyn matrix~\cite{Ka}.
In this case $\sigma(s,t)$ can be computed explicitly as the Legendre transform of the free energy $f(H,V)$
of the six-vertex model on a torus (with $\Delta=0$) in the presence of electric fields $H$ and $V$:
$$
\sigma(s,t)=\min_{H,V} \left( Hs + Vt - f(H,V) \right).
$$
The function $f(H,V)$ is given by the double integral
\[
f(H,V)=\frac{1}{(2\pi i)^2}\int_{|z|=\exp(H)}\int_{|w|=\exp(V)} \ln|P(z,w)| \frac{dz}{z}\frac{dw}{w},
\]
where
\begin{equation}
\label{SpectralCurve}
P(z,w)=a(wz-1)+b(z+w)
\end{equation}
is the spectral polynomial of the Kasteleyn matrix, see \cite{Ka,McCoyWu,KOS}. 
Note that $f(H,V)$ is convex, $\text{det}(\partial_{i}\partial_{j} f)>0$, and $\sigma(s,t)$ is concave, $\text{det}(\partial_{i}\partial_{j} \sigma)<0$.

Euler-Lagrange equations for the large deviation rate functional (\ref{LDf}) can be written explicitly as follows (see \cite{KOS,KO} for details).
Consider complex valued 
functions $z(x,y)$ and $w(x,y)$ such that 
\begin{equation}\label{arg}
arg(z(x,y))=\pi \pa_x h(x,y), \ \ arg(w(x,y))=-\pi \pa_y h(x,y).
\end{equation}
Then the Euler-Lagrange equations for $h(x,y)$ can be written as a system of 
equations for $z(x,y)$ and $w(x,y)$ as
\begin{equation}\label{DEhf}
\pa_y \log (z)+\pa_x \log(w)=0, \ \  P(z,w)=0.
\end{equation}

\begin{figure}[t!]
\begin{center}
\includegraphics[width=1.0\textwidth, angle=0.0, scale=0.6]{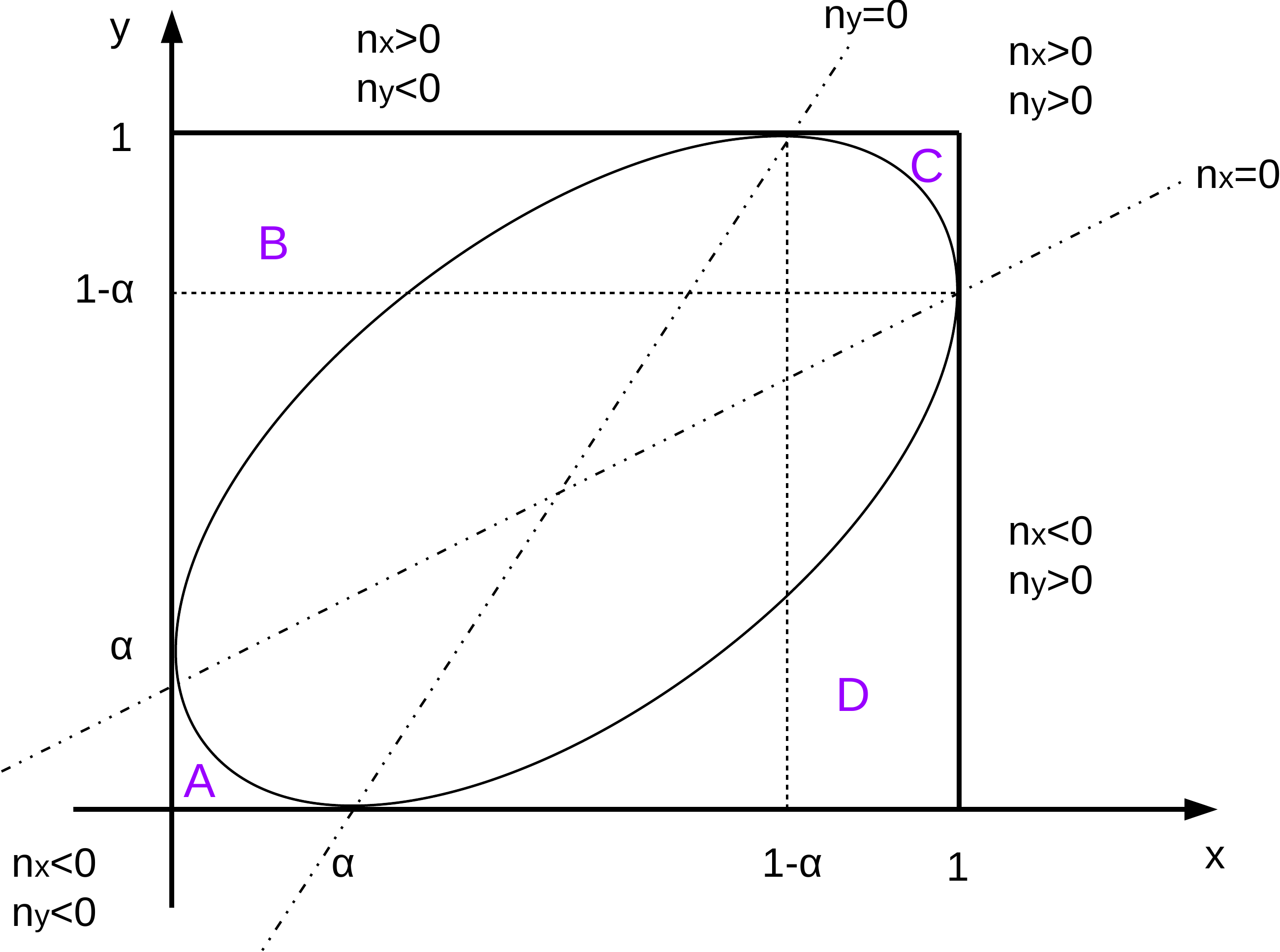}
\end{center}
\caption{\label{Ell-1} The $xy$ plane with the denoted domains. The ellipse is the boundary of the limit shape. The height function smooth
inside the ellipse and linear outside. Here, we also show the lines $n_x=0$, $n_y=0$ and areas $A$, $B$, $C$, and $D$. }
\end{figure}

\subsection{The limit shape}
Here we describe the limit shape height function $h_0(x,y)$ for DW boundary 
conditions. We use the result of \cite{Pronko} where the density of the horizontal edges 
occupied by the paths is derived.

Define the function $D(x,y)$ as
$$
D(x,y) = \alpha (1-\alpha) \left[ \frac{(y-x)^{2}}{\alpha}+\frac{(1-x-y)^{2}}{1-\alpha}-1 \right].
$$
Here, the parameter $\alpha$ is determined by values of Boltzmann weights of the model as
$$
\alpha=\frac{b}{a}.
$$
In Baxter's parametrization $ \alpha=\tan{(u)}$.
Define the region $E= \{ (x,y) | D(x,y) \le 0 \}$ as the interior of the ellipse  $\partial E= \{ (x,y) | D(x,y)= 0 \}$
which is inscribed  in the square $0\leq x \leq 1, 0\leq y \leq 1$ as it is shown in Fig. \ref{Ell-1}. The 
ellipse is the boundary of the limit shape, or the ``arctic curve''~\cite{Pronko2011}.

The following expression was derived in \cite{Pronko}: 
\begin{equation}\label{DyHF}
\partial_{y} h_0(x,y) = \left\{
  \begin{array}{cc}
    \frac{1}{\pi} \text{arccot} \left( \frac{-n_{y}}{\sqrt{-D(x,y)}} \right), &  (x,y) \in E \\
    0, & (x,y) \in A \cup B, \enskip n_{y}<0 \\
    1, & (x,y) \in C \cup D, \enskip n_{y}>0 \\
  \end{array}
\right.
\end{equation}
Regions $A,B,C,D$ are shown in Fig. \ref{Ell-1}
Here $n_{y}=(1-\alpha)(y-x)+\alpha (1-x-y)=x+(2\alpha-1) y -\alpha$.

Integrating this expression, we obtain the following formula for 
the limit shape height function itself:
\begin{equation}\label{HF}
h_0(x,y) = \left\{
  \begin{array}{cc}
    \frac{1}{\pi} \left( y \, \text{arccot} \left[ \frac{-n_{y}}{\sqrt{-D(x,y)}} \right] -\frac{1}{2} \, \text{arctan}\left[\frac{-x^2+(y-\alpha)(\alpha-1)+x(1+y-2y\alpha)}{(1-x-\alpha)\sqrt{-D(x,y)}} \right] + \right. & \\
 \left. +(\frac{1}{2}-x) \, \text{arctan} \left[ \frac{-n_{x}}{\sqrt{-D(x,y)}} \right] \right) +\frac{x}{2} & (x,y) \in E \mbox{ }\& \mbox{ } x<1-\alpha \\
\frac{1}{\pi} \left( y \, \text{arccot} \left[ \frac{-n_{y}}{\sqrt{-D(x,y)}} \right] -\frac{1}{2} \, \text{arctan}\left[\frac{-x^2+(y-\alpha)(\alpha-1)+x(1+y-2y\alpha)}{(1-x-\alpha)\sqrt{-D(x,y)}} \right] + \right. & \\
 \left. +(\frac{1}{2}-x) \, \text{arctan} \left[ \frac{-n_{x}}{\sqrt{-D(x,y)}} \right] \right) +\frac{x}{2}-\frac{1}{2} & (x,y) \in E \mbox{ }\& \mbox{ } x \ge 1-\alpha \\
    0 & (x,y) \in A \\
    x & (x,y) \in B \\
    x+y-1 & (x,y) \in C \\
    y & (x,y) \in D
  \end{array}
\right.
\end{equation}
Here $n_{x}=(1-\alpha)(y-x)+\alpha (x+y-1)=y+(2\alpha-1) x -\alpha$.
\begin{figure}[b!]
\begin{center}
\vspace{1mm}\includegraphics[width=0.4\textwidth, angle=0.0]{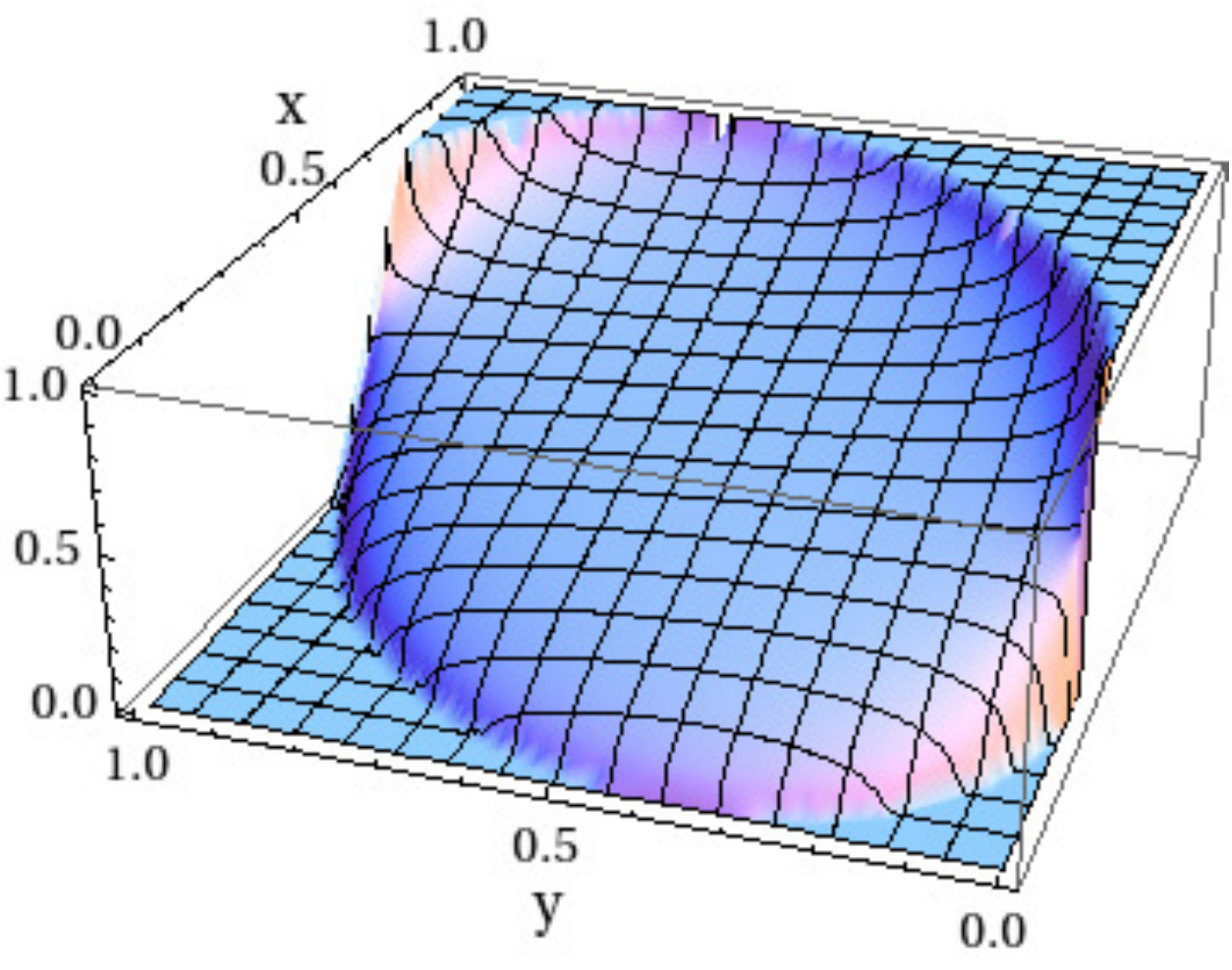}
\end{center}
\caption{\label{f1} The density of horizontal edges occupied by paths, or $\partial_{y} h_0(x,y)$, for $\alpha=9/25$.}
\end{figure}

Differentiating this expression in $x$, we obtain the density of edges occupied with horizontal paths:
\begin{equation}
\label{partialX}
\partial_{x} h_0(x,y) = \left\{
  \begin{array}{cc}
    -\frac{1}{\pi} \text{arctan} \left( \frac{-n_{x}}{\sqrt{-D(x,y)}} \right)+\frac{1}{2}, &  (x,y) \in E \\
    0, & (x,y) \in A \cup D, \enskip n_{x}<0 \\
    1, & (x,y) \in B \cup C, \enskip n_{x}>0 \\
  \end{array}
\right.
\end{equation}

Here we use the branch of the function $\text{arctan}$ which behaves as
\begin{equation}
-\frac{1}{\pi} \text{arctan} \left( \frac{-n_{x}}{\sqrt{-D(x,y)}} \right) \to  \left\{
  \begin{array}{cc}
    -\frac{1}{2}, & n_{x}<0, \enskip (x,y) \in A \cup D \\
    \frac{1}{2}, & n_{x}>0, \enskip (x,y) \in B \cup C \\
  \end{array}
\right.
\end{equation}
when $(x,y)$ approach to the boundary of $E$.

\begin{figure}[t!]
\begin{center}
\vspace{1mm}\includegraphics[width=0.4\textwidth, angle=0.0]{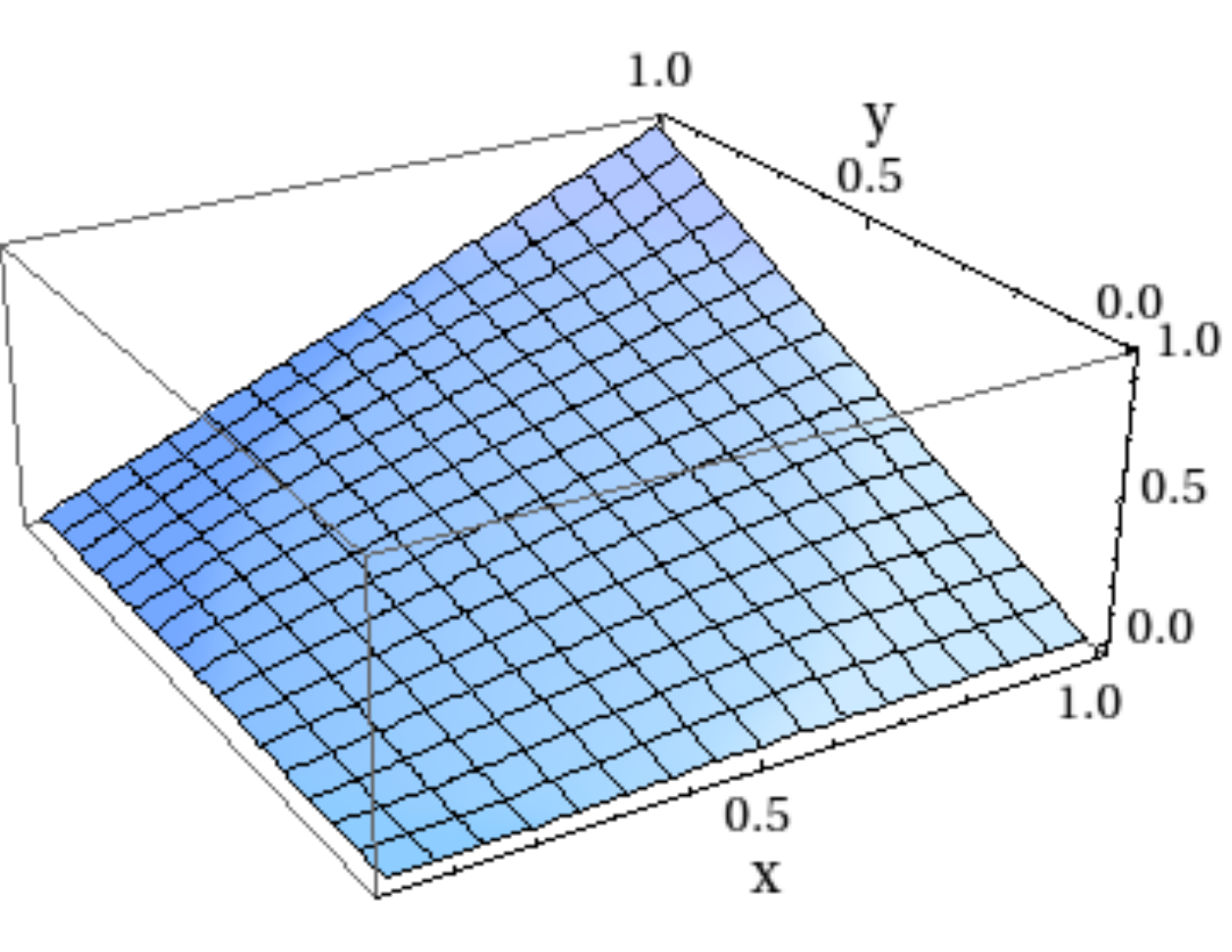}
\end{center}
\caption{\label{f2} The limit shape height function $h_{0}(x,y)$, Eq.~(\ref{HF}), for $\alpha=9/25$.}
\end{figure}

As an example, in Fig.~\ref{f1} we show the partial derivative of the height function~(\ref{HF}) for $\alpha=9/25$.
Inside the arctic curve, it is given by the nontrivial part of Eq.~(\ref{DyHF}) and outside that one, it equals to zero or one. 
The limit shape height function $h_{0}(x,y)$, Eq.~(\ref{HF}), is shown in Fig.~\ref{f2}.
%

\subsection{The function $z(x,y)$}
An important property of functions $z(x,y)$ and $w(x,y)$ is that $z$ maps 
the inner part of the ellipse $D(x,y)=0$ (the arctic curve) to the upper half plane.
Indeed, as we saw in the previous section, the $x$-derivative of the height function~(\ref{partialX})
is non-negative when $0\leq x\leq 1$ and, therefore, the imaginary part of the function $z(x,y)$
is also non-negative.

In our case we already know the height function, so in order to find functions $z$ and $w$
it is sufficient to solve the algebraic equation in (\ref{DEhf}). 
Moreover, since we know the height function, we know the arguments of $z$ and $w$, therefore, we just have to solve
the equation $P(z,w)=0$ for absolute values of $z$ and $w$. This will give 
the conformal mapping $z$ from the interior $E$ of our ellipse to the 
upper half of the complex plane.

Solving the quadratic equation $P(z,w)=0$ for the absolute values of $z$ and $w$ and taking into account (\ref{arg}),
we obtain:
\begin{eqnarray}\label{zw}
\nonumber |z|=\frac{1}{2 \, a \, b \, \sin[-\pi \pa_yh_0]}\Bigg(a^2 \sin[ \pi (\pa_xh_0-\pa_yh_0) ]+b^2 \sin[ \pi (\pa_xh_0+\pa_yh_0)] \\
\hspace{1cm}\mp \sqrt{4 a^2 b^2 (\sin[-\pi \pa_yh_0])^2+\bigg(a^2 \sin[\pi (\pa_xh_0-\pa_yh_0)]+b^2 \sin[\pi (\pa_xh_0+\pa_yh_0)]\bigg)^2}\Bigg),\label{z} \\
\nonumber |w|=\frac{1}{2 \, a \, b \, \sin[\pi \pa_xh_{0}]}\Bigg(a^2 \sin[\pi (\pa_xh_0-\pa_yh_0)]+b^2 \sin[\pi (\pa_xh_0+\pa_yh_0)] \\
\hspace{1cm}\pm \sqrt{4 a^2 b^2 (\sin[-\pi \pa_yh_{0}])^2+\bigg(a^2 \sin[ \pi (\pa_xh_0-\pa_yh_0)]+b^2 \sin[ \pi (\pa_xh_0+\pa_yh_0)])\bigg)^2}\Bigg) \label{w}.
\end{eqnarray}

We almost constructed the mapping $z: E \to H=\{ z | Re(z) \geq 0 \},  \pa E \to \RR$. The last step is to determine the signs in (\ref{z}).
In the Appendix~\ref{A1}, we determine the signs and the mapping.
It maps the boundary of $E$ bijectively to the real line in the following way:
\begin{itemize}
\item $z: A\cap \pa E\to (0, \frac{a}{b})$, $A\cap B\mapsto 0$, $A\cap D\mapsto \frac{a}{b}$
\item $z: B\cap \pa E\to (-\frac{b}{a}, 0)$, $B\cap C\mapsto -\frac{b}{a}$, $B\cap A\mapsto 0$
\item $z: C\cap \pa E\to (-\infty, -\frac{b}{a})$, $C\cap D|_{C}\mapsto -\infty$, $C\cap B\mapsto -\frac{b}{a}$
\item $z: D\cap \pa E\to (\frac{a}{b},\infty)$, $D\cap A\mapsto \frac{a}{b}$, $D\cap C|_{D}\mapsto \infty$
\end{itemize}

\subsection{The two-point correlation function}
In the continuum limit, the fluctuations of the height function are described
by the massless Euclidean quantum Bose field in the interior of the arctic curve with the metric determined
by the second variation $S^{(2)}$ of the large deviation rate functional (\ref{LDf}) computed at the limit shape height function.
It reads
$$
S^{(2)}[h_{0}]= \frac{1}{2} \iint_{\mathcal D} \left( \partial_{1}^{2} \sigma(\vec{\nabla} h_{0}) (\partial_{x} \phi)^{2}+
2 \partial_{1} \partial_{2} \sigma(\vec{\nabla} h_{0}) \partial_{x} \phi \, \partial_{y} \phi +
\partial_{2}^{2} \sigma(\vec{\nabla} h_{0}) (\partial_{y} \phi)^{2} \right) \, dx \, dy.
$$

The mapping $z$ brings the functional $S^{(2)}$ with the kernel defined on functions on the interior of $E$ to 
the Dirichlet functional for the Laplace operator acting on functions on the upper half of the complex plane. 
This defines the two-point correlation function for fluctuations of the height function on $E$
as the Green's function for the Laplace operator on the upper half plane with Dirichlet boundary 
conditions on the real line:
\begin{equation}
\label{Corr2Pi}
\left \langle \phi(x_{1},y_{1}),\phi(x_{2},y_{2}) \right \rangle = -\frac{1}{2\pi} \ln \Bigg| \frac{z(x_{1},y_{1})-z(x_{2},y_{2})}{z(x_{1},y_{1})- \overline{z(x_{2},y_{2})}} \Bigg|.
\end{equation}
Here, $\phi$ is the fluctuation field from (\ref{eq1chi}).
The formula (\ref{Corr2Pi}) means that the two-point correlation function at the free fermionic point~($\Delta=0$) has a logarithmic dependence on the distance between points $(x_{1},y_{1})$ and $(x_{2},y_{2})$, when this distance is small~\cite{Kenyon}.

For free fermionic models, local correlation functions (multipoint correlation functions) $\langle \phi(\vec{r}_{1}),\ldots,\phi(\vec{r}_{n}) \rangle$ of fluctuations of the height function are determined by the two-point correlation functions through the Wick's formula.

%
\begin{figure}[t!]
\begin{center}
\vspace{1mm}\includegraphics[width=1.0\textwidth, angle=0.0]{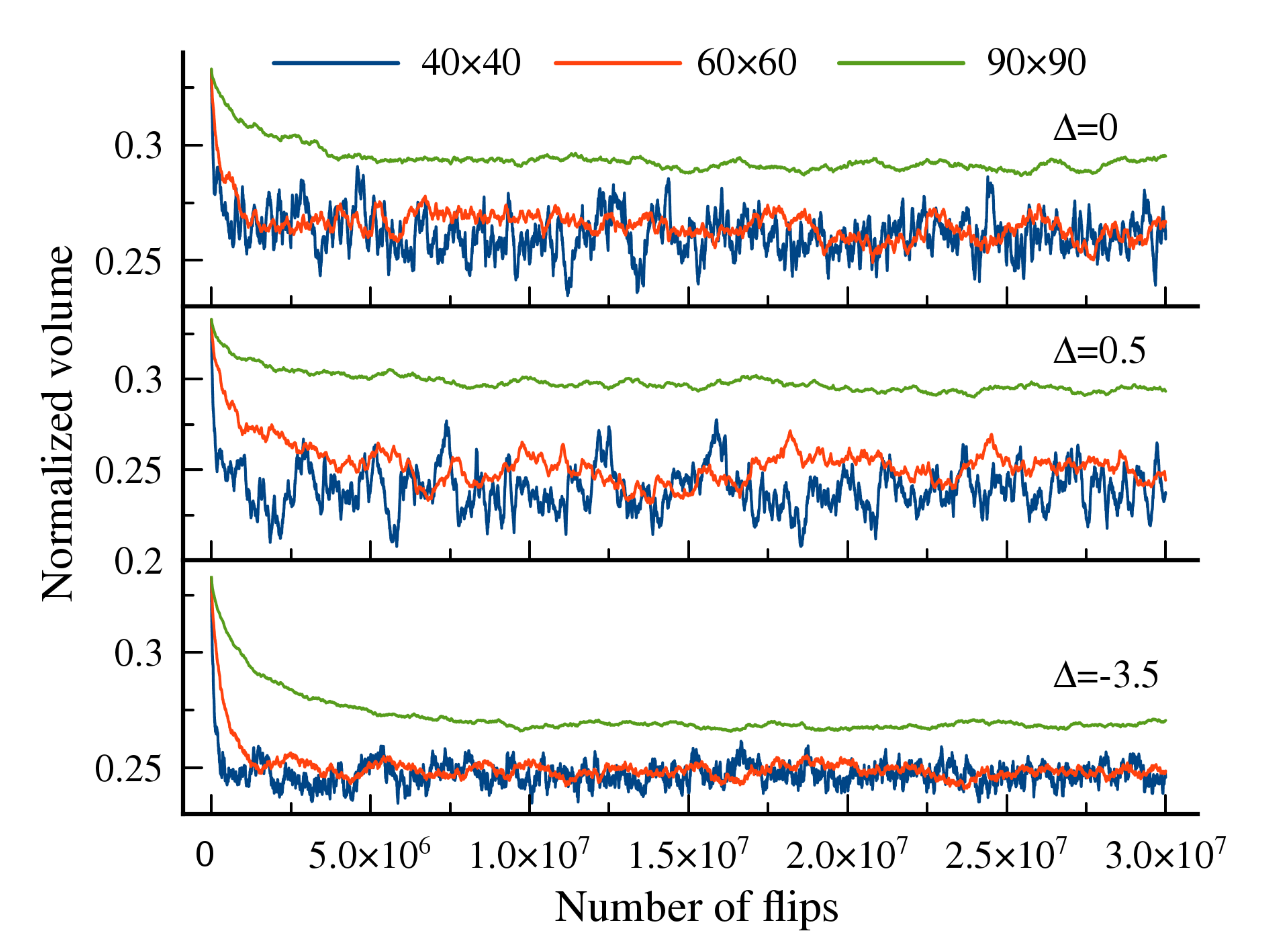}
\end{center}
\caption{\label{thermalizationFig} The behavior of the normalized volume under the height function values as a function of a number of flips for three different lattice sizes and three values of $\Delta$: $0$, $1/2$, and $-7/2$ of the six-vertex model.}
\end{figure}
%

\section{Numerical results}\label{num}

\subsection{Computation of observables and the thermalization.}

As it was mentioned in the introduction, we use the Markov chain sampling algorithm to generate a sequence of random states of the six-vertex model~\cite{R2005}.
This method is known as the Markov chain Monte-Carlo simulation.
It is based on the special choice of the transition probabilities to transfer from an arbitrary distribution to the desired one.
It is also known as the Metropolis algorithm~\cite{Metropolis}.
An overview of these numerical methods and their applications to statistical mechanics can be found in~\cite{LandauBinder2014}.
See~\cite{Zvonarev,AS,Keesman,Korepin1,Korepin2} for related numerical simulations.

The idea of Markov sampling is to create a random process that will follow the most likely states in the model.
This is guaranteed by the choice of the matrix of transition probabilities which is symmetrizable (detailed balanced condition) by the diagonal matrix with entries given by Boltzmann weights of the system.
This condition (plus an assumption of nondegeneracy of the largest eigenvalue) 
also guarantees the asymptotical convergence of the process to the Boltzmann distribution 
starting from any distribution. Also, in this case the Boltzmann distribution is the Perron-Frobenius 
eigenvector of the matrix of transition probabilities\footnote{These are all standard facts about Markov processes, for details see
for example \cite{LandauBinder2014,Seneta,Markov}.}.

When a random process is constructed, the expectation values of observables with respect to the 
Boltzmann distribution can be computed by averaging along the random process. 
This procedure is especially effective when the Boltzmann distribution is concentrated in a small neighborhood of
the most likely state (the limit shape). In probability theory this is known as large deviations, in non-equilibrium 
statistical physics this is known as a hydrodynamic limit. 

In dimer models it was proven rigorously~\cite{CKP} 
that there exists a most probable state and the probability for any other state to be ``macroscopically distant'' from
it is exponentially suppressed:
\begin{equation}\label{localization}
Prob(h)\propto \exp\left[N^2(S[h_{0}]-S[h])\right].
\end{equation}
Here, $h_0$ is the height function corresponding to the limit shape~(\ref{HF}). It minimizes the large deviation rate functional.
The minimal value $S[h_{0}]$ is exactly (minus) the free energy of the system. 

The six-vertex model with $\Delta=0$ is equivalent to a dimer model.
Therefore, in this case we can use the probability distribution and results from the corresponding dimer model.
For other values of $\Delta$ in the six-vertex model the analysis is more complicated, but we expect a similar 
structure of the distribution, suggesting the formation of the limit shape $h_{0}$.

The localization (concentration) of random states near the limit shape makes the numerical computation of observables
easy once the Markov process is thermalized i.e. when it moves along the states in a vicinity of the limit shape.
Thus, the main challenge for computing observables is to know when the process is thermalized.
Unfortunately, it is very hard to have an effective criterium for thermalization.
Instead, we use a simple empirical technique: we monitor the fluctuations of the normalized volume under the height
function
\[
vol(h)=\frac{1}{N^3}\sum_{(n,m)} h(n,m).
\]
As it is clear from Fig.~\ref{thermalizationFig}, the normalized volume ``drifts'', when the process is not yet thermalized.
Then it starts to fluctuate around the normalized volume under the limit shape $h_{0}$.
Thus, we can start measurements to compute observables using the Markov chain simulations.
For example, Fig.~\ref{thermalizationFig} shows that for the lattice of size $90\times 90$ and $\Delta=1/2$
it is safe to start averaging after about $10^{7}$ flips~\cite{R2005}.
\begin{figure}[t!]
\begin{center}
\begin{tabular}{cc}
\begin{minipage}{0.5\linewidth}
\includegraphics[width=1.0\textwidth, angle=0.0]{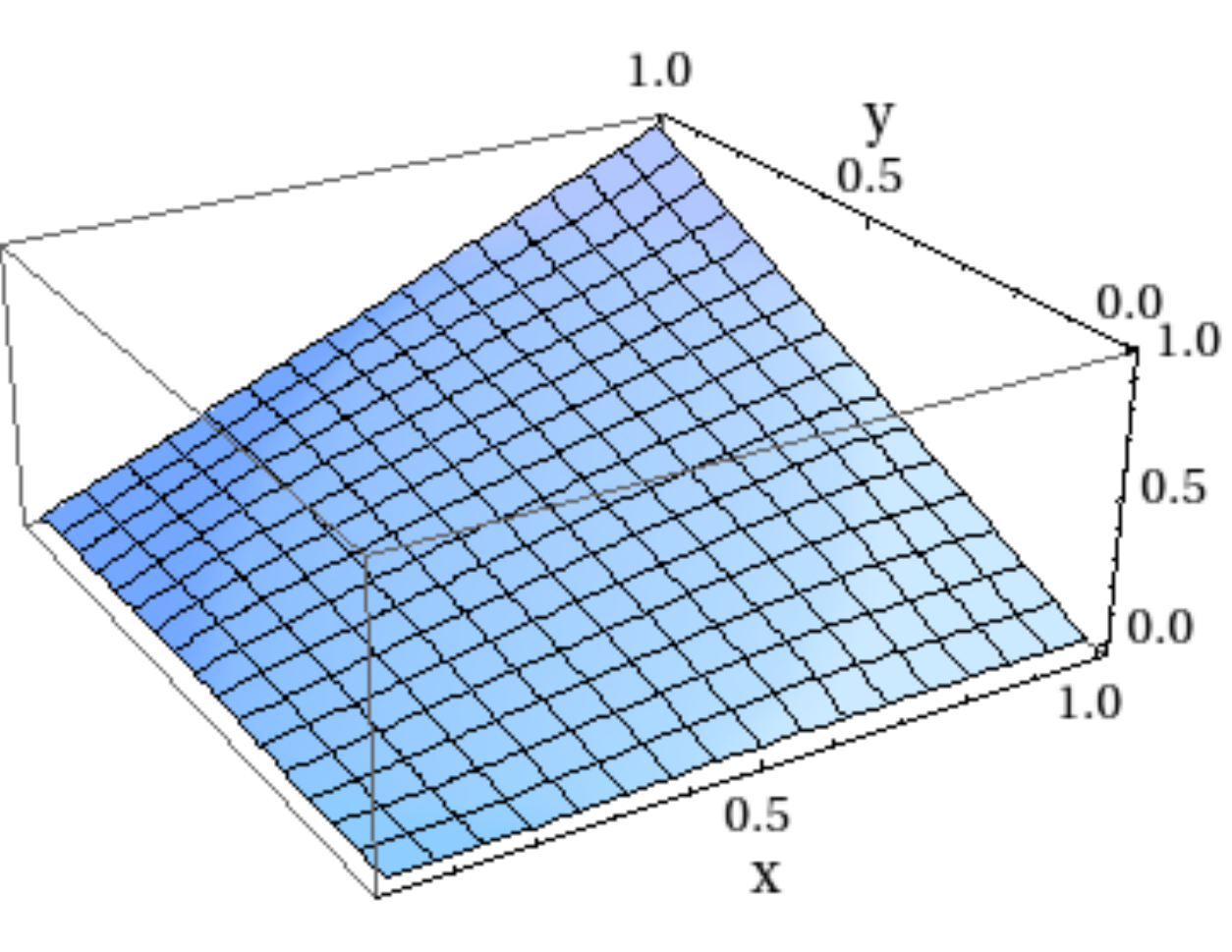}
\end{minipage}
&
\begin{minipage}{0.5\linewidth}
\includegraphics[width=1.0\textwidth, angle=0.0]{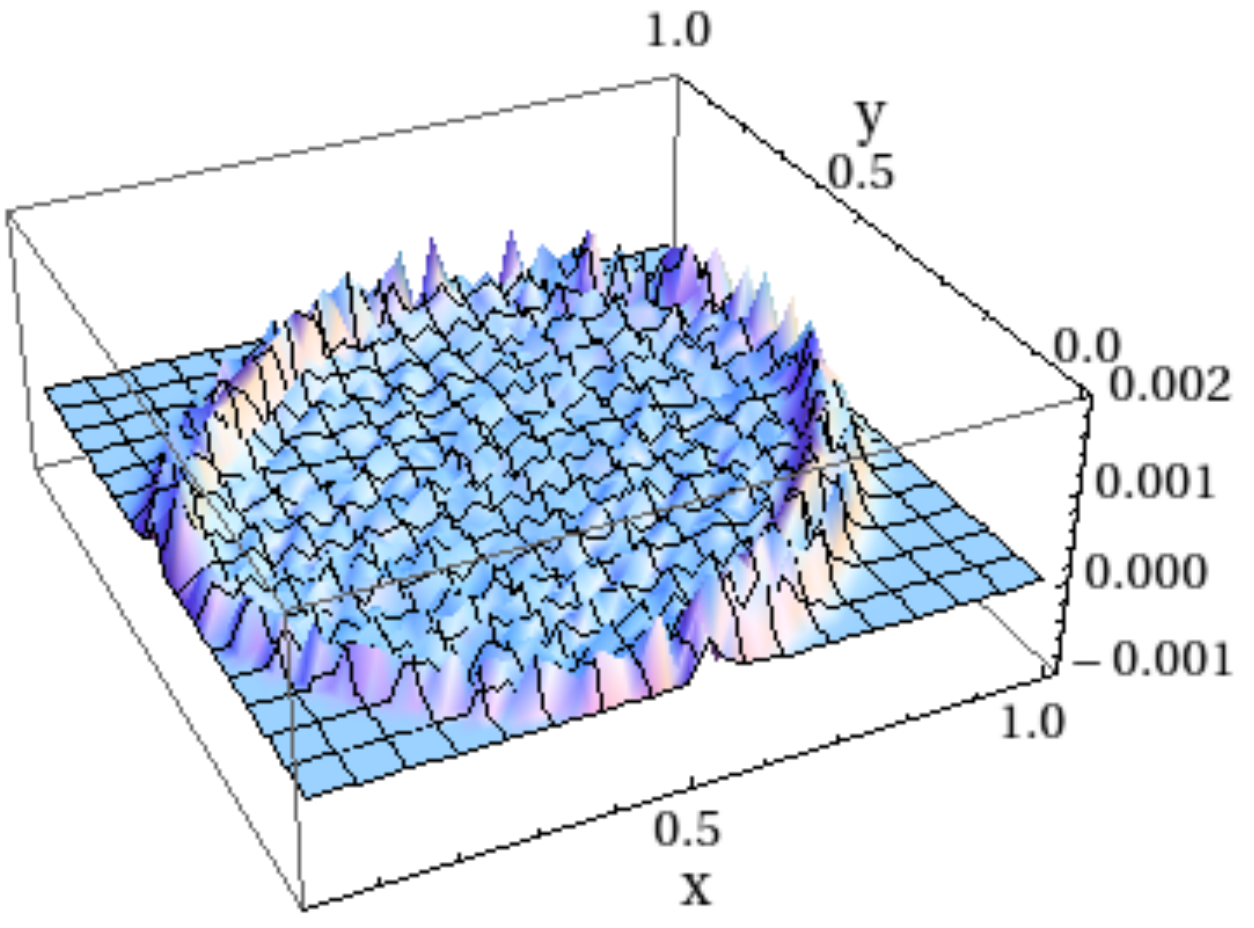}
\end{minipage}
\end{tabular}
\end{center}
\caption{\label{FigCompare4040} (\textit{left plot}) The numerical height function for $\Delta=0$ and for the lattice size of $40\times 40$. The parameter $\alpha=9/25$. The height function is the result of averaging over measurements. (\textit{right plot}) The difference between the theoretical limit shape height function and the one obtained from numerical simulation.}
\end{figure}

Once the thermalization is achieved, we compute an observable by time averaging:
\begin{equation}\label{O}
\left \langle O\right \rangle \simeq \frac{O(s_1)+\dots +O(s_K)}{K}.
\end{equation}
Here $s_i$  is a random state at time $T_i$ counting from the first measurement, $K$ is the total number of measurements.
The right side depends on random states $s_i$ and is a random variable, but as $K\to \infty$
it converges to the Boltzmann expectation value.
Of course, numerically $K\to \infty$ simply means large values.
We will use this to compute the limit shape $h_{0}$
and correlation functions.
In particular, the two-point correlation function of points $(x_{i},y_{i})$ and $(x_{j},y_{j})$ is calculated as
\begin{equation}\label{ncf}
\left \langle\phi(x_{i},y_{i}),\phi(x_{j},y_{j})\right \rangle = \left \langle \chi(x_{i},y_{i}) \chi(x_{j},y_{j}) \right \rangle-\left \langle \chi(x_{i},y_{i})\right \rangle \left \langle \chi(x_{j},y_{j}) \right \rangle,
\end{equation}
where 
\begin{equation}\label{nhf}
\left \langle \chi(x_{i},y_{i})\right \rangle = \frac{1}{K}\sum_{k=1}^K \chi_k(x_{i},y_{i}),
\end{equation}
the height function $\chi$ is from Eq.~(\ref{eq1chi}),
and indices $i,j=1,\ldots,N$ numerate the lattice sites.
Here $\chi_{k}$ are random variables, but the sum represents a deterministic quantity as $K\to\infty$.
\begin{figure}[t!]
\begin{center}
\begin{tabular}{cc}
\begin{minipage}{0.5\linewidth}
\includegraphics[width=1.0\textwidth, angle=0.0]{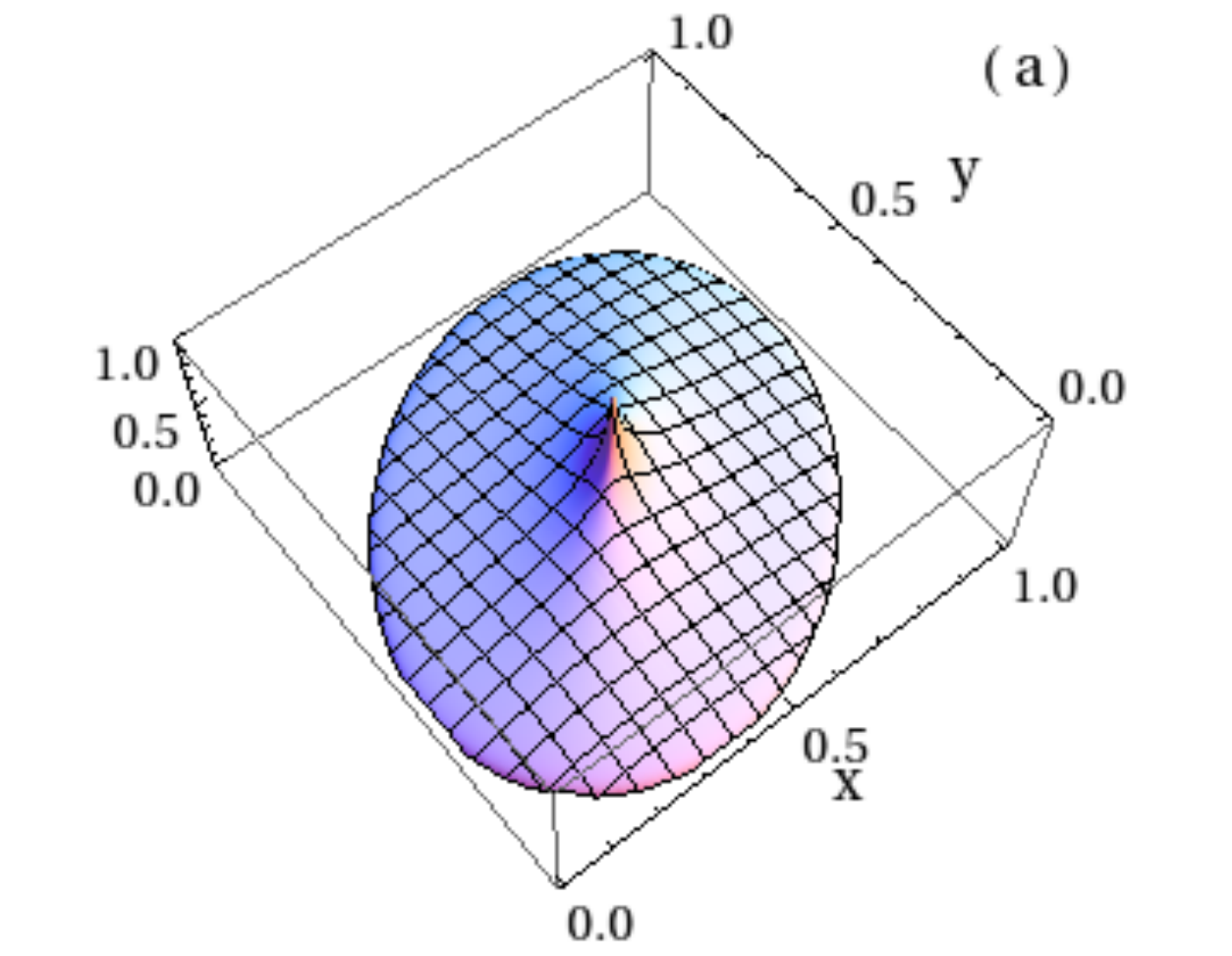}
\end{minipage}
&
\begin{minipage}{0.5\linewidth}
\includegraphics[width=1.0\textwidth, angle=0.0]{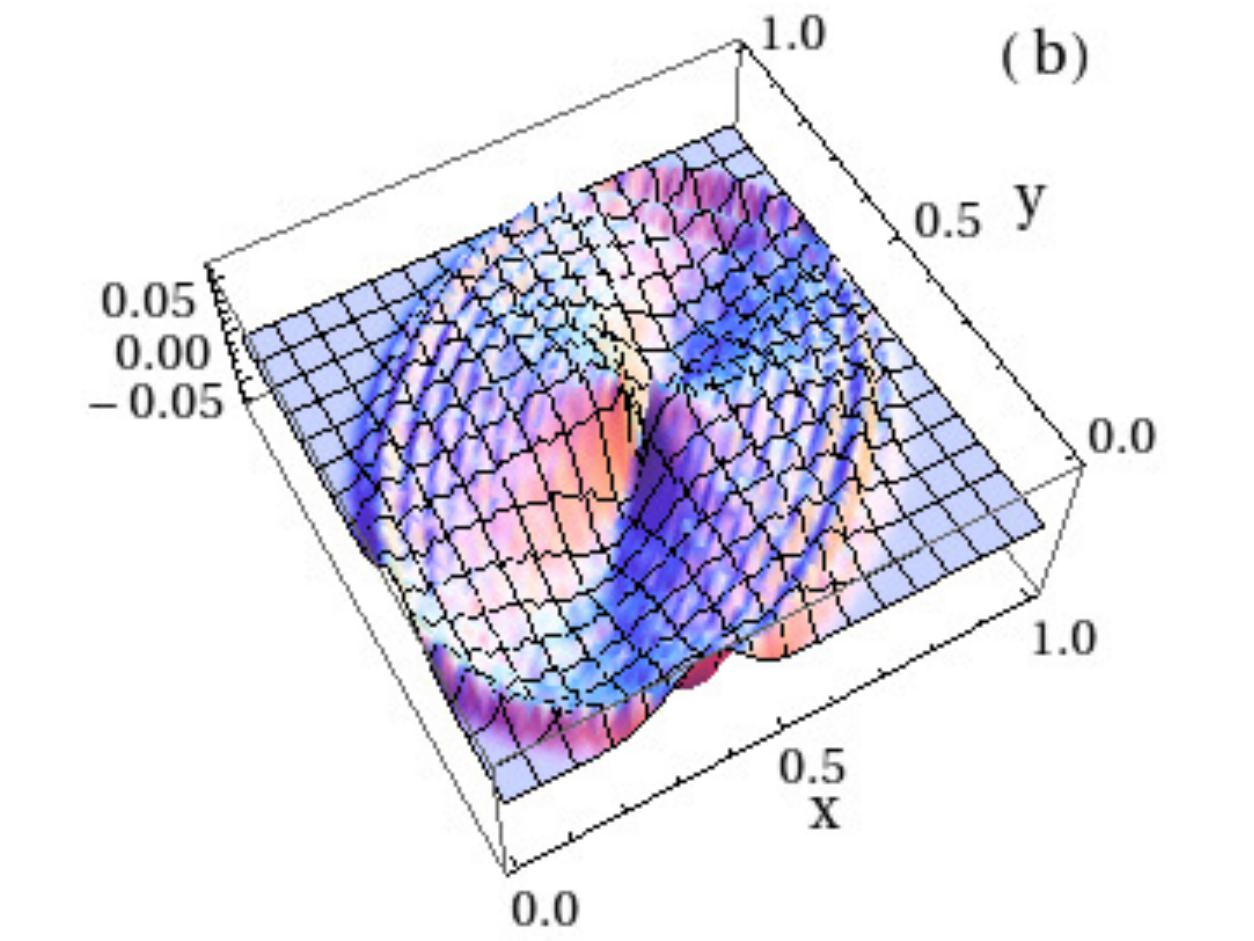}
\end{minipage} \\
\begin{minipage}{0.5\linewidth}
\includegraphics[width=1.0\textwidth, angle=0.0]{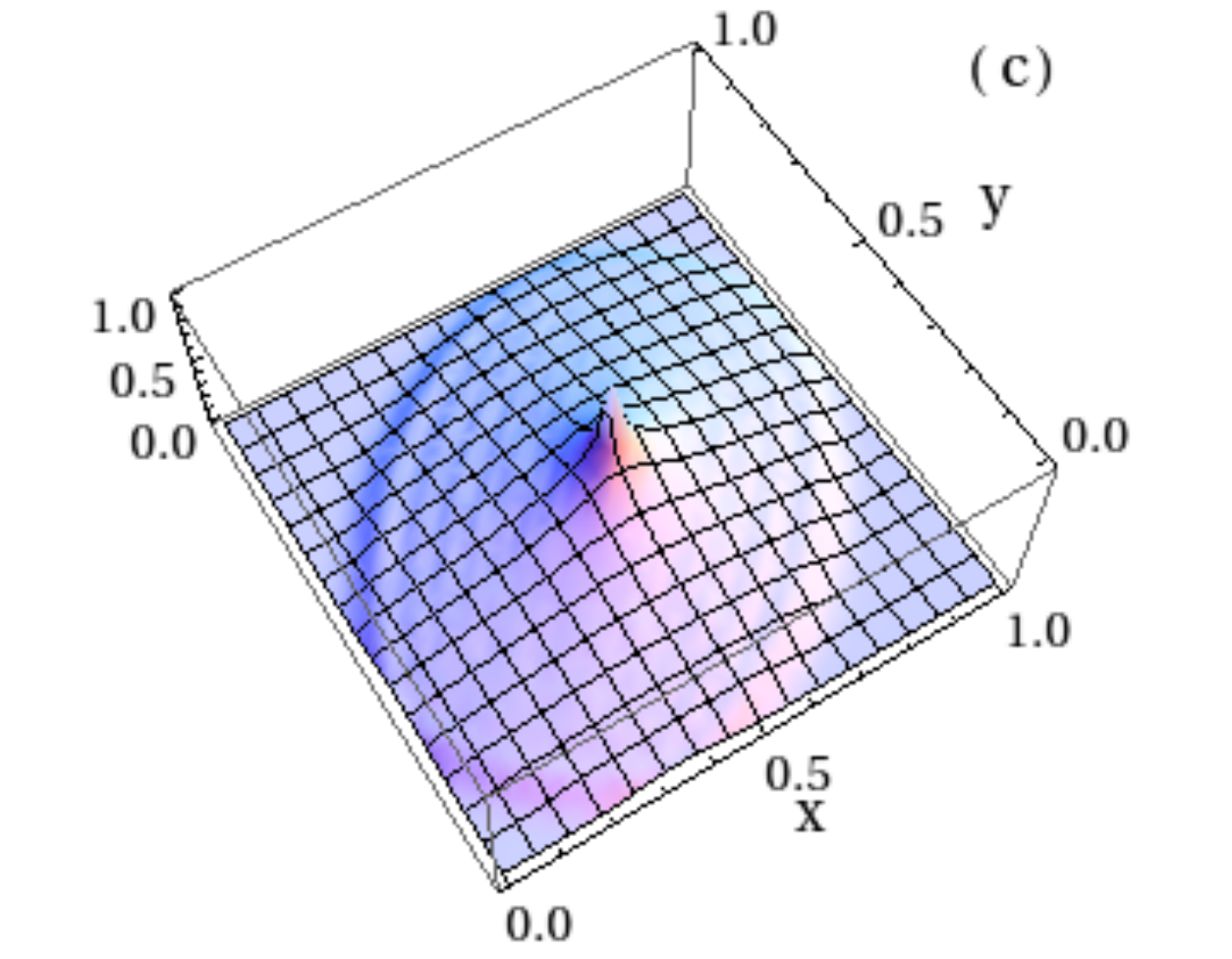}
\end{minipage}
&
\begin{minipage}{0.5\linewidth}
\includegraphics[width=1.0\textwidth, angle=0.0]{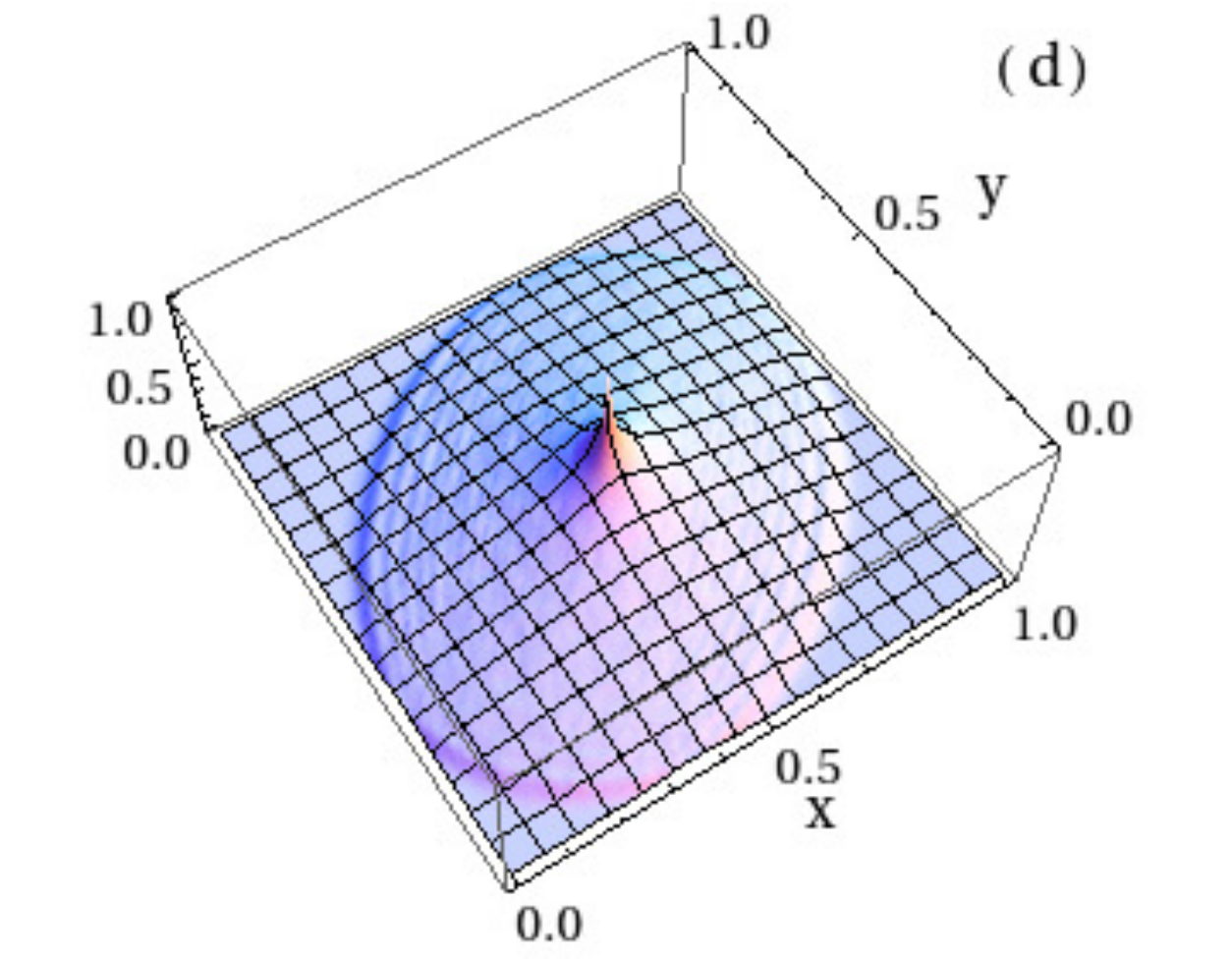}
\end{minipage}
\end{tabular}
\end{center}
\caption{\label{CorrCompareTheor406090} The plots of (a) the exact limit shape two-point correlation function inside the arctic curve for $\Delta=0$, (b) the difference between the numerical correlation function on the square lattice of $60\times 60$ and the exact one, the numerical ones on the square lattices of sizes (c) $40\times 40$ and (d) $90\times 90$. The parameter $\alpha=9/25$.}
\end{figure}

\subsection{Numerical computation of the limit shape and correlation functions at the free fermionic point.}
We start by comparison of the calculated height function with the exact one, the limit shape $h_{0}$ given by Eq.~(\ref{HF}), for $\Delta=0$ and $\alpha=9/25$.
The difference between the exact height function and the numerical one is shown in Fig.~\ref{FigCompare4040}.
It should be noted that the numerical height function is smooth since it was averaged over a number of measurements, as described above.
The difference between $h_{0}$ and the numerical height function reveals the Airy asymptotic near the boundary of the limit shape.
The difference vanishes as the lattice size $N\to \infty$.
\begin{figure}[t!]
\begin{center}
\vspace{1mm}\includegraphics[width=1.0\textwidth, angle=0.0]{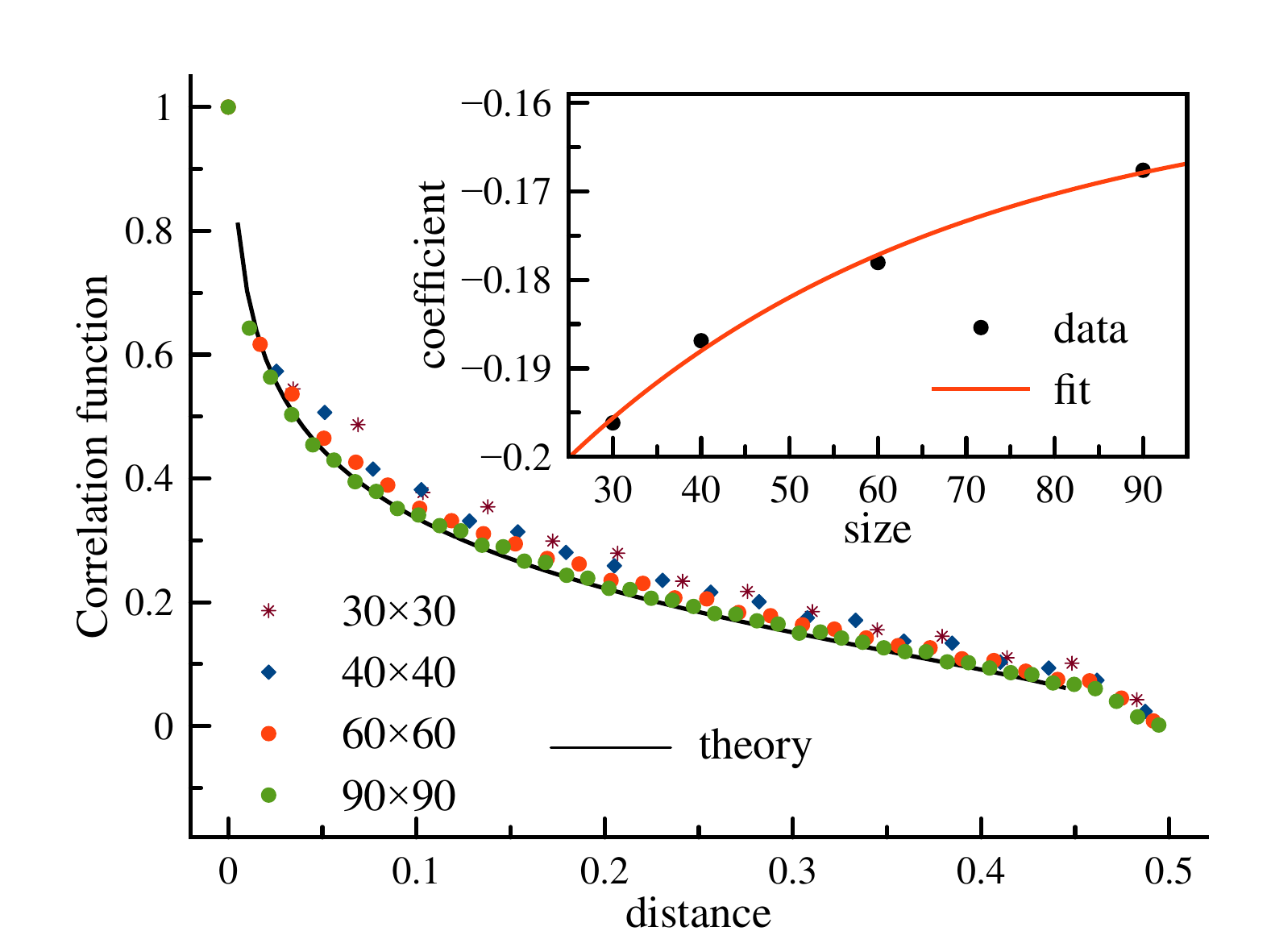}
\end{center}
\caption{\label{DetailedCorrCompare6060} The values of the two-point correlation function along the slice $y=0.5$ for $\Delta=0$. The curve shows the theoretical limit shape correlation function~(\ref{Corr2Pi}). The points are the calculated values for different lattice sizes. The inset shows the values of numerically obtained coefficient against logarithm in Eq.~(\ref{Corr2Pi}) with respect to the lattice size as well as a fit of these coefficients.}
\end{figure}

The results of computations of two-point correlation functions $\left \langle\phi(x_{i},y_{i}),\phi(x_{j},y_{j})\right \rangle$ are presented in Fig.~\ref{CorrCompareTheor406090}.
We show plots of data~(\ref{ncf}) for one point $(x_i,y_i)=(0.5,0.5)$ and another point $(x_j,y_j)$ running through the square $[0,1]\times [0,1]$ of the lattice domain with step $1/N$.
The plot (a) shows the theoretical limit shape correlation function, given by Eq.~(\ref{Corr2Pi}).
The plots (c),(d) represent the computed values itself
obtained in a ``single run'' of the Markov process described above for different linear sizes $N$ of the system.
The measurements are taken after thermalization after each several hundred thousand iterations of the process.
The total number of measurements is $10^{5}$.
The plot (b) shows the difference between the theoretical exact values and the numerical computation.
Again, the difference shows the Airy waves propagating from the boundary of the limit shape.
As in the case of the height function, one can see the Airy waves which decrease with increasing of $N$.

The agreement of theoretical values of the correlation function and the corresponding  numerical values can
be seen qualitatively by comparing  pictures in Fig.~\ref{CorrCompareTheor406090}.
When $\Delta=0$ the six-vertex model maps to a dimer model and therefore in the limit $N\to \infty$
correlation functions converge to conformally invariant correlation functions (\ref{Corr2Pi}).
One can see the logarithmic behavior in the two-point correlation function
in the vicinity of $(x_i,y_i)$, where $\left \langle\phi(\vec{r_{i}}),\phi(\vec{r}_{j})\right \rangle \sim -1/(2\pi)\, \ln{|\vec{r}_{i}-\vec{r}_{j}|}$.

By plotting the results of numerics along the slice $y=0.5$ we can examine the logarithmic behavior carefully, see Fig.~\ref{DetailedCorrCompare6060}.
There is a good agreement between the theoretical result and numerical data for different lattice sizes:
the calculated values converge to the theoretical prediction as the lattice size increases.
The numerical values of the coefficient against the logarithm in Eq.~(\ref{Corr2Pi}) have been obtained from the fit to data.
They are shown in the inset of Fig.~\ref{DetailedCorrCompare6060}.
We see that, as the lattice size increases, $N\to \infty$, the value of the coefficient against the logarithm tends to the exact one $-1/(2\pi)\approx -0.159155$.
For example, a fit by logarithm for the lattice size $90\times 90$ yields the coefficient $-0.167$.
A fit of the values of the coefficient (red line in the inset), in turn, gives the approximate value for the infinite lattice to be $-0.1584\pm0.0082$, which is close to the exact one.
%

\subsection{Numerical results for $\Delta=1/2$}

The agreement of theoretical and numerical results at the free fermionic point, $\Delta=0$, suggests that the numerics should work equally well for other values of $\Delta$, where the analytical results are still unknown.
Here, we present numerical results for $\Delta=1/2$.
We choose this value of $\Delta$ randomly, but note that it is also known as the combinatorial point where the model has many extra interesting features~\cite{ZJdisser}.

\begin{figure}[t!]
\begin{center}
\begin{tabular}{ccc}
\begin{minipage}{0.33\linewidth}
\includegraphics[width=1.0\textwidth, angle=0.0]{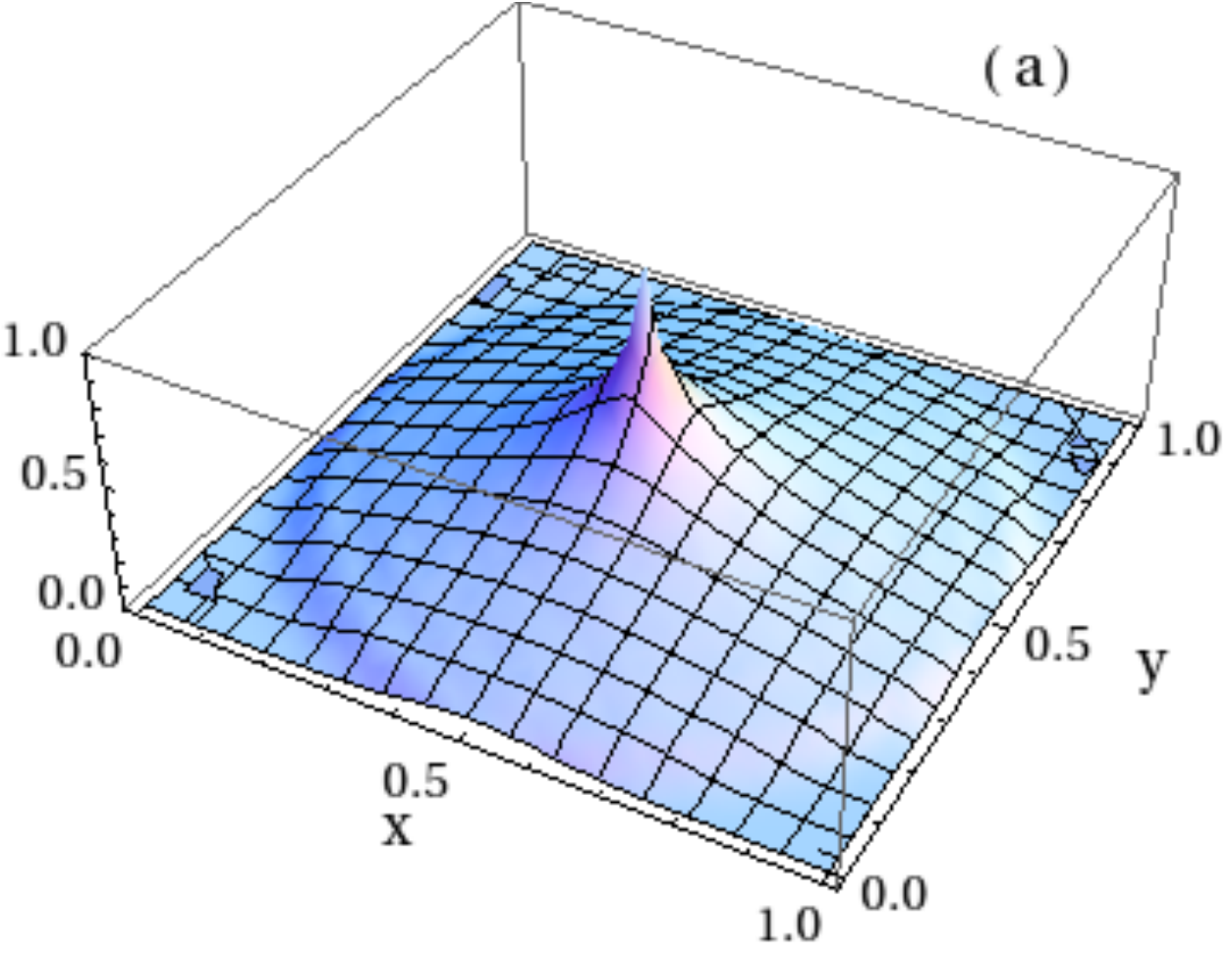}
\end{minipage}
&
\begin{minipage}{0.33\linewidth}
\includegraphics[width=1.0\textwidth, angle=0.0]{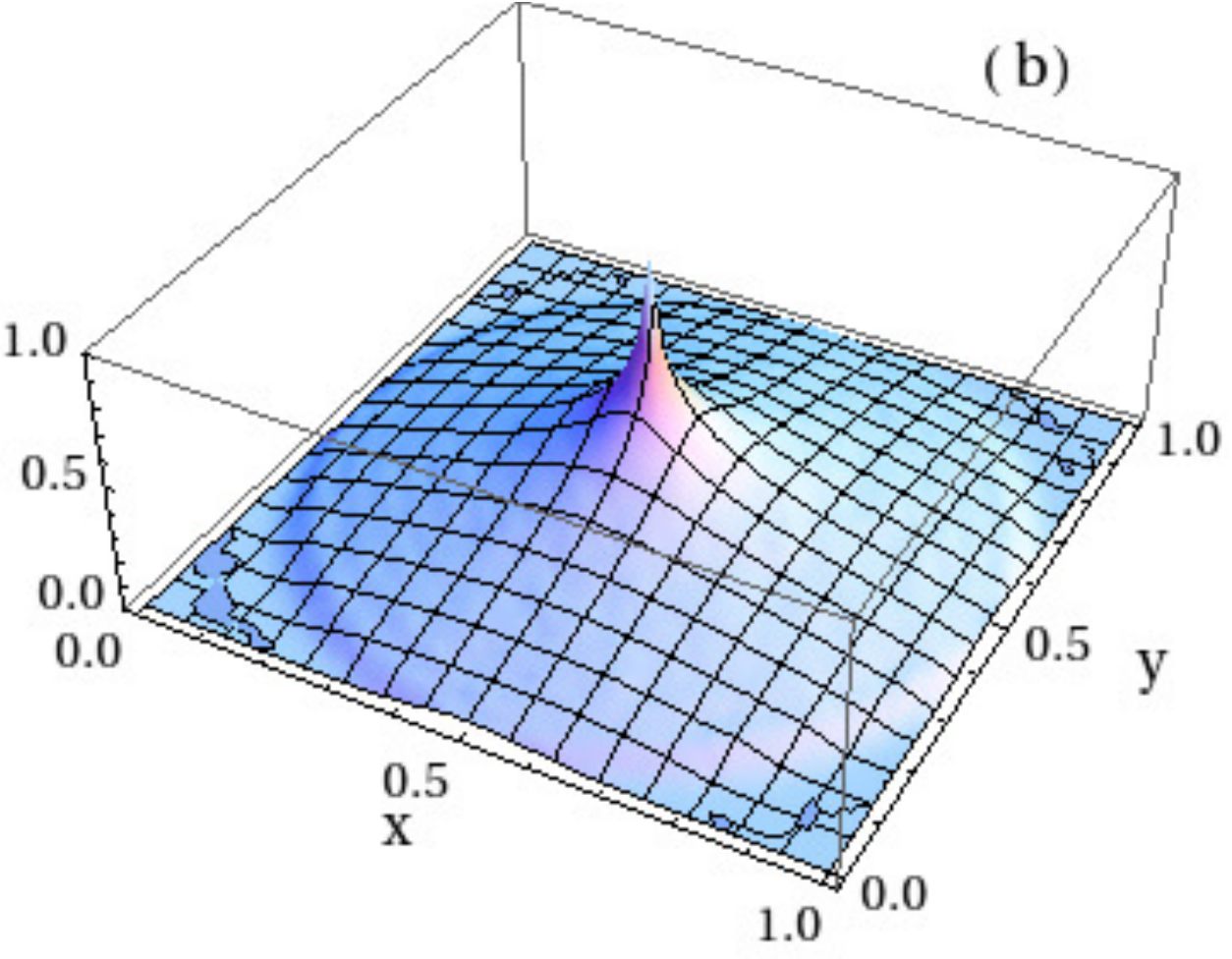}
\end{minipage}
&
\begin{minipage}{0.33\linewidth}
\includegraphics[width=1.0\textwidth, angle=0.0]{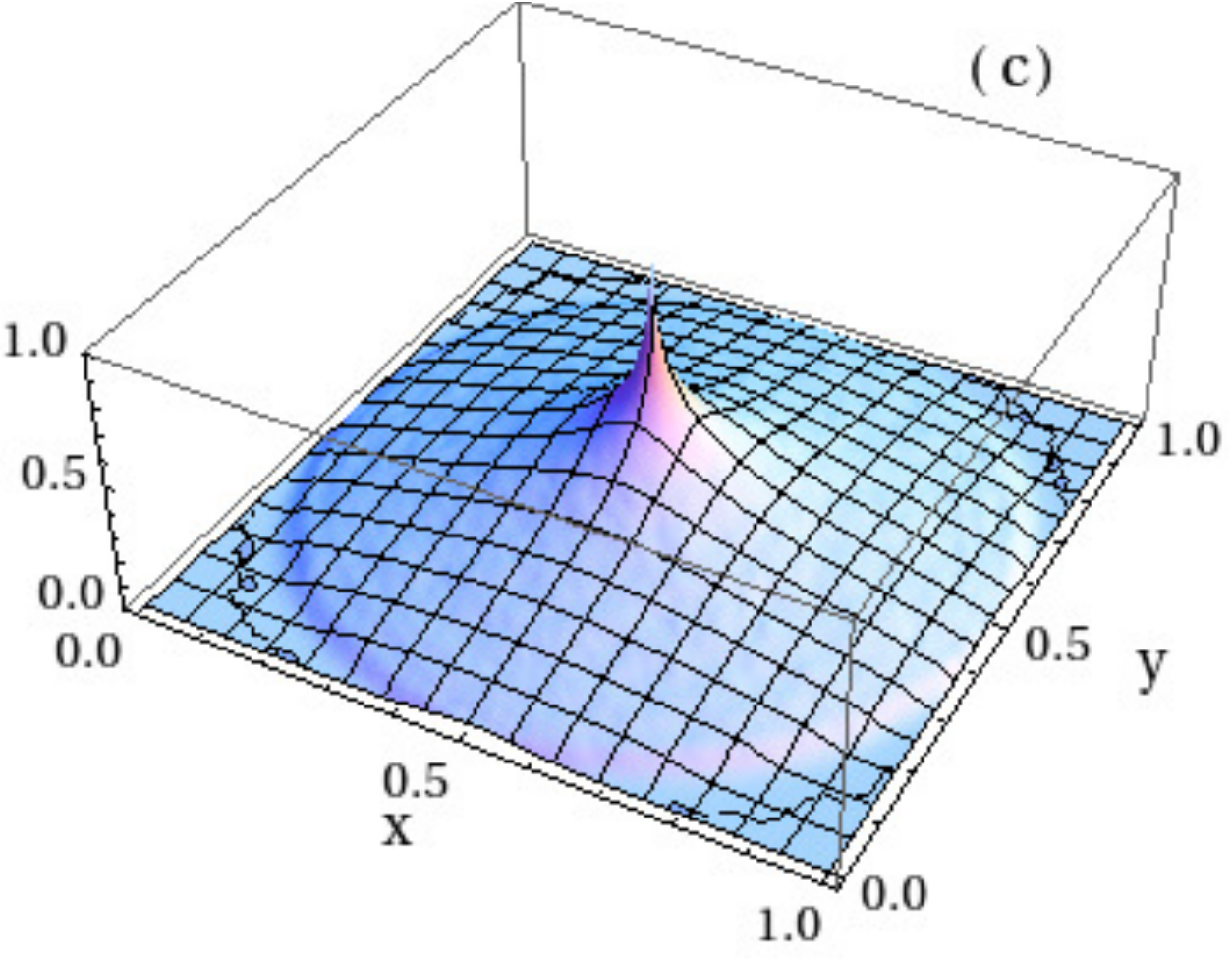}
\end{minipage}
\end{tabular}
\end{center}

\caption{\label{f1Delta05} The correlation function for the case $\Delta=1/2$, $a/c=b/c=1$. The shown data are for lattice sizes (a) $40\times 40$, (b) $60\times 60$, and (c) $90\times 90$.}
\end{figure}

The results of numerical computation of the two-point correlation function for $a/c=b/c=1$
are shown in Fig.~\ref{f1Delta05}.
Three plots correspond to three lattice sizes: $40\times 40$, $60\times 60$, and $90\times 90$.
The behavior of the correlation function at short distances, as expected, is very similar to that for the free fermionic point in Fig.~\ref{CorrCompareTheor406090}.

\begin{figure}[t!]
\begin{center}
\vspace{1mm}\includegraphics[width=1.0\textwidth, angle=0.0]{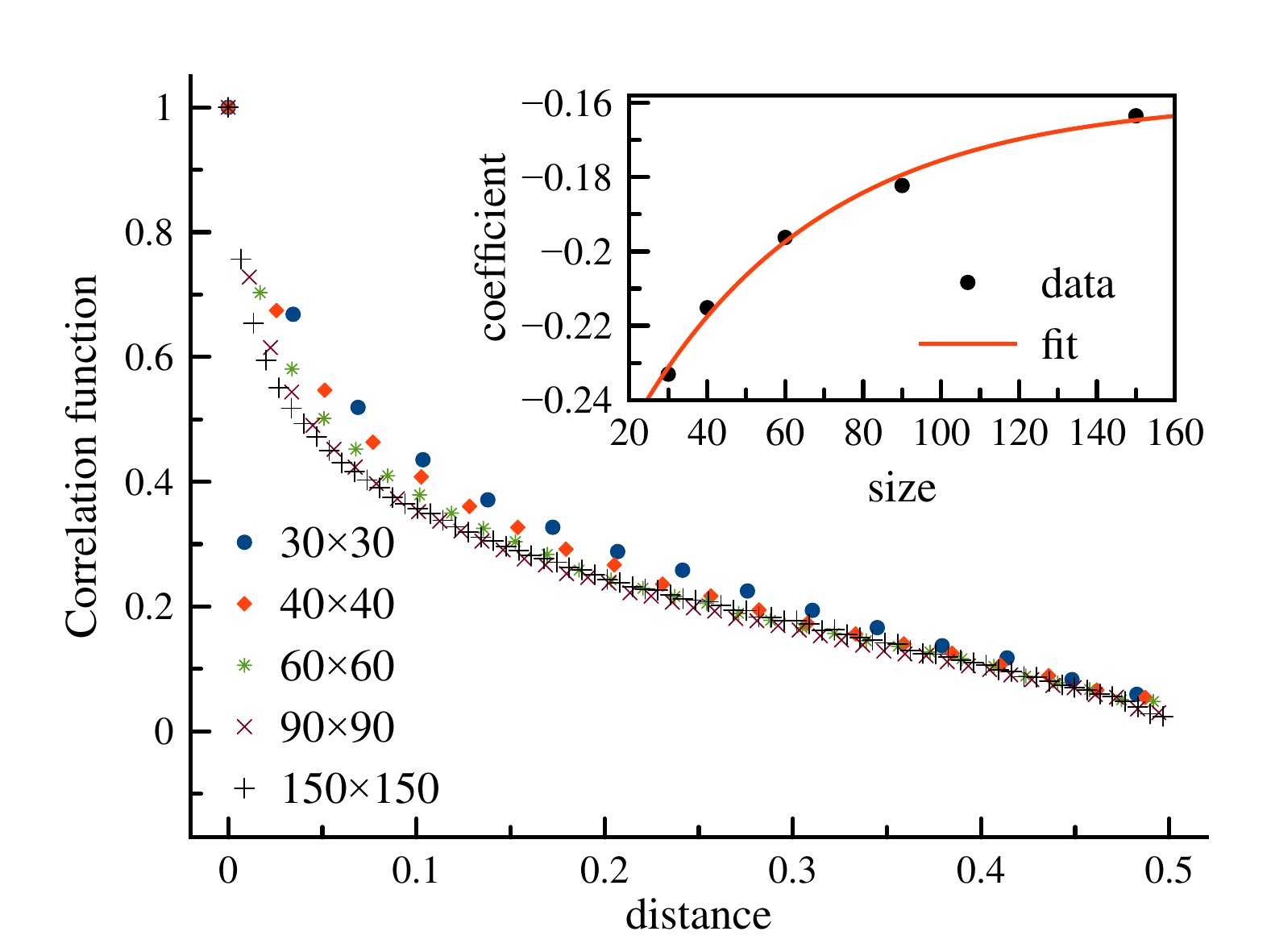}
\end{center}
\caption{\label{f2Delta05}The values of the two-point correlation function along the slice $y=0.5$ for $\Delta=1/2$. The calculated values for different lattice sizes are shown. The inset shows the values of numerically obtained coefficient against logarithm in the logarithm-like fit with respect to the lattice size as well as a fit of these coefficients.}
\end{figure}

The numerical values of the two-point correlation function along the slice $y=0.5$ for $\Delta=1/2$ are shown in Fig.~\ref{f2Delta05}.
The short distance asymptotic of the correlation function is again logarithmic. This is 
in agreement with the fact that the model is in the disordered phase. The difference with the 
free fermionic case is that the global correlation function is not given by a conformal 
mapping anymore, but is given by an effective Gaussian field theory, see for example the discussion in~\cite{GBDJ}.
However, as in any disordered phase, at distances which are larger than the lattice step, but much smaller than the characteristic size of the lattice, the correlation functions are still given by an effective conformal field theory.
In the case of the six-vertex model, this is $c=1$ Gaussian CFT model with logarithmic correlators.

The fitted coefficient against logarithm in Eq.~(\ref{Corr2Pi}) approaches the exact value $-1/(2\pi)$ as the lattice size $N \to \infty$ in this case as well.
However, the numerical values of this coefficient when $\Delta=1/2$ are notably worse than the same values for $\Delta=0$. For $\Delta=1/2$, the values for smaller lattices are systematically smaller than those for $\Delta=0$.
They are naturally expected to converge to the exact value, but the rate of a convergence is less than for $\Delta=0$.
For example, when the lattice size is $90\times 90$, the fit by logarithm gives the value $-0.182$ for the coefficient against $\log$.
The convergence is shown in the inset of Fig.~\ref{f2Delta05} with the extrapolated value for the infinite lattice being $-0.1588\pm0.0058$, which is still close to the expected $-1/(2\pi)$.

As in the  case $\Delta=0$ one can see the Airy waves near the boundary of the limit shape. They disappear when $N$ is increasing.

\begin{figure}[t!]
\begin{center}
\vspace{1mm}\includegraphics[width=0.750\textwidth, angle=0.0]{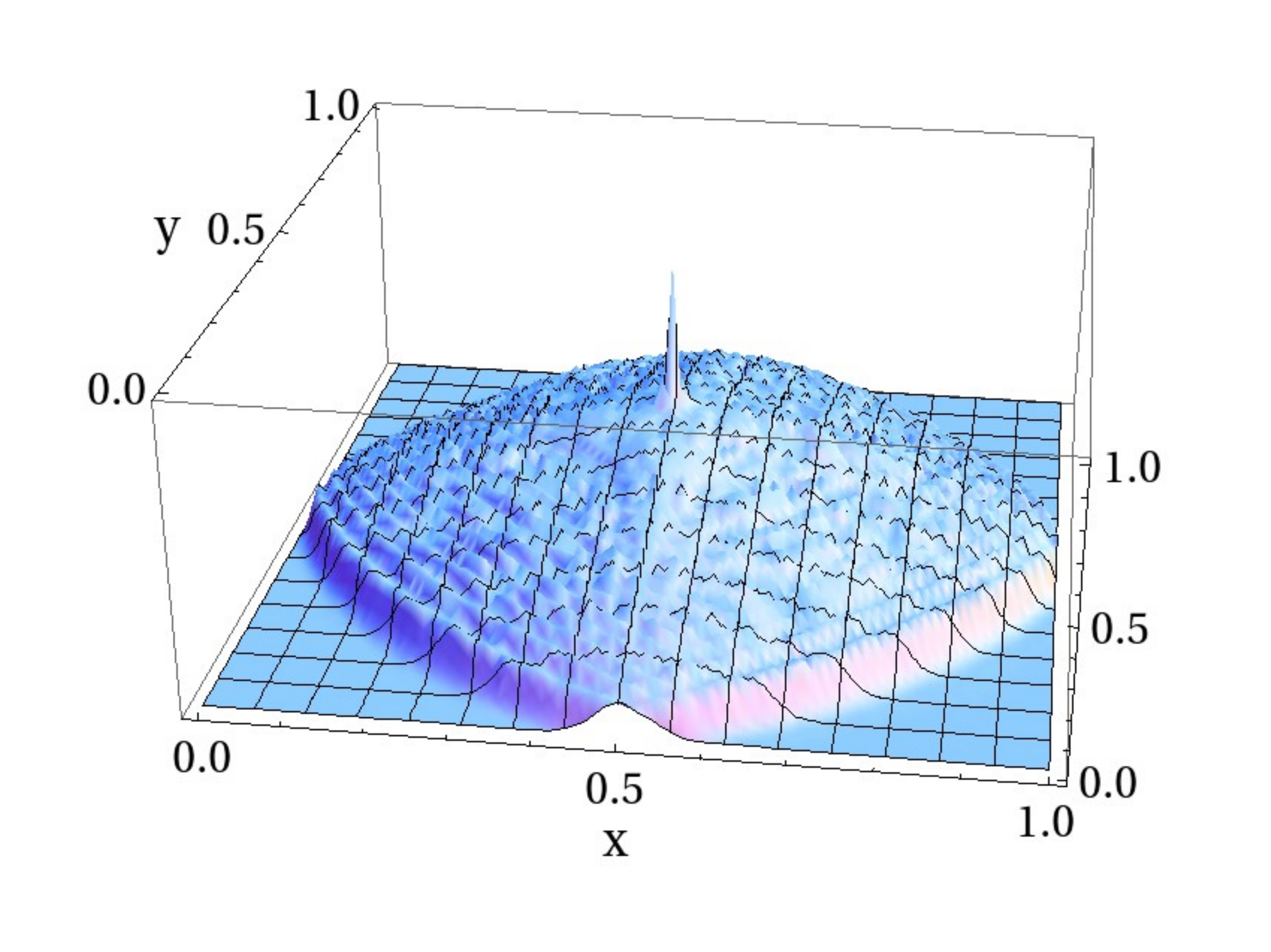}
\end{center}
\caption{\label{af}The values of the two-point correlation function for the antiferroelectric phase $\Delta=-7/2$. The lattice size is $N=90$. The sharp peak is given by the exponential fall which is fitted as $0.484+0.516 \exp{\left((-99\pm 4) |\vec{r}_{i}-\vec{r}_{j}|\right)}$.}
\end{figure}

\subsection{Numerical results for $\Delta=-7/2$}

For $\Delta<-1$, the antiferroelectric phase of the six-vertex model is opened in the form of a diamond shape droplet.
This droplet has already been observed earlier~\cite{R2005,Zvonarev,Cugliandolo,Korepin1}.
The Gaussian field $\phi(x,y)$ in this region is massive which predicts the exponential decay of correlation functions at short distances.
The numerical observation of the exponential decay of correlation function in this phase is challenging. 
In order to carry out such computations of correlation functions at distances deep in the antiferroelectric droplet, 
the characteristic length of the droplet should be much larger than the correlation length. But for these values of $N$ and $\Delta$ 
the thermalization is expected exponentially long~\cite{Ra}.

We carried out calculations of two-point correlation functions for $\Delta=-7/2$ and the lattice size $N=90$.
The result is given in Fig.~\ref{af}, and the thermalization was presented in Fig.~\ref{thermalizationFig}.
The numerics are in qualitative agreement with the theoretical prediction that the correlation function should exponentially decrease.
One can see a sharp peak over a relatively flat background.
The sharp peak is given by the exponential fall and can be fitted as $0.484+0.516 \exp{\left((-99\pm 4) |\vec{r}_{i}-\vec{r}_{j}|\right)}$.

The background ``pillow'' in Fig.~\ref{af} is expected to be a result of ``mesoscopic'' effects. The lattice size $N=90$ is relatively small and 
correlation functions get affected by the Airy processes on the boundaries of the disordered region. 
The parallel GPU computations on large lattices may resolve this issue~\cite{AS}.
More careful analysis of comparative values of the linear size of the droplet and of the correlation length
will be given in a separate publication both numerically and from the exact solution.

\section{Conclusion}
In this paper, we numerically calculated the two-point correlation functions for the six-vertex model with the domain wall boundary conditions.
The disordered ($|\Delta|<1$) phase has mainly been studied.
Particular attention was paid to the free fermionic point ($\Delta=0$), for which the correlation function has been also obtained analytically in the thermodynamic limit, $N\to\infty$.
The logarithm-like behavior of correlation functions at the small scales has been confirmed.
For antiferroelectric phase, the exponential decrease of the correlator has been observed.
The numerics for $N=90$ and $\Delta=-7/2$ show that it might be interesting to study correlation functions in the mesoscopic region where the size of the antiferroelectric droplet is comparable to the correlation length in the antiferroelectric phase.
We plan to continue studies of correlation functions and, in particular, their asymptotics in the limit $\Delta \to -1-0$
when the relatively small characteristic size of the droplet requires computations on large lattices.
For such lattices, the implementation of Markov sampling on GPU may be of great practical significance.

\appendix

\section{The symmetry}\label{Sym}

Consider the following mapping of the weights and configurations of the six-vertex model.
On weights it acts as $a\mapsto b, \ \ b\mapsto a,\ \ c\mapsto c$. On states, it replaces 
each horizontal edge which is not occupied by a path with an edge occupied by a path
and each occupied edge by an empty edge. It is clear that the probability measure is
invariant with respect to this mapping:
\[
Prob_{a,b,c}(\gamma, \beta)=Prob_{b,a,c}(\overline{\gamma}, \overline{\beta})
\]
Here $\gamma$ is a state of the six-vertex model and $\beta$ is the boundary configuration of paths, that is
fixed. For the DW boundary conditions we have
\[
Prob_{a,b,c}(\gamma)=Prob_{b,a,c}(\overline{\gamma})
\]
Note that the probability measure depends only on the ratios $a:b:c$, therefore, when $c\neq 0$, we can set $c=1$.
At the free fermionic point $\Delta=0$, this, together with the symmetry described above, implies that the probability measure with $\alpha=b/a$ is equal to the one with $\alpha^{-1}$. Therefore, we can assume that $0<\alpha<1$.

\section{Derivation of signs}\label{A1}

\begin{figure}[t!]
\begin{center}
\includegraphics[width=1.0\textwidth, angle=0.0, scale=0.6]{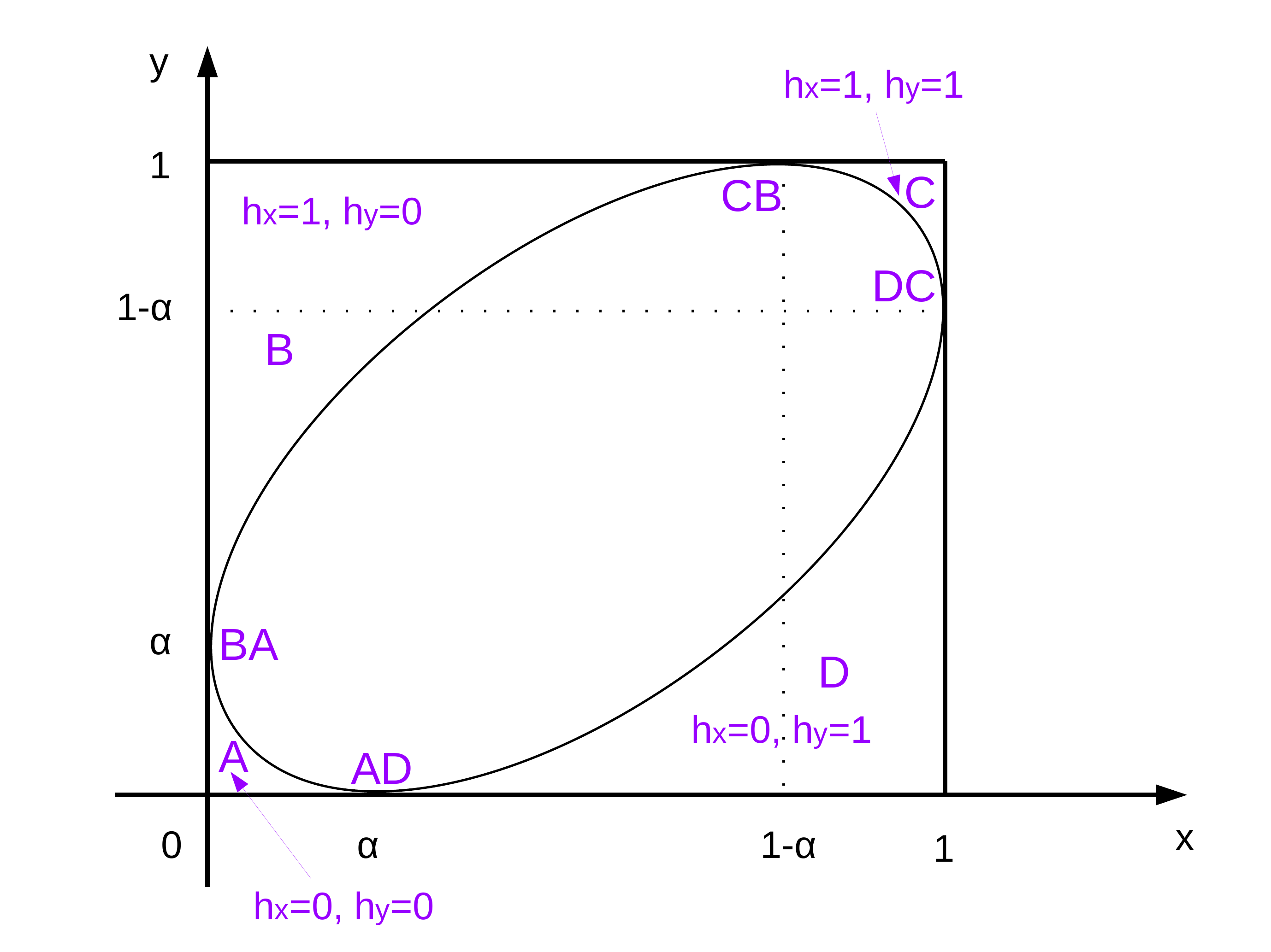}
\end{center}
\caption{\label{figAppendix1} The limit shape height function has a constant slope in regions $A$,$B$,$C$,$D$ (frozen regions)
and it is strictly convex inside the ellipse. At points $DC$, $CB$, $BA$, $AD$ one of the slopes
of the limit shape is discontinuous along $\pa E$. The partial derivatives of the limit shape height function $h_{x}\equiv \partial_{x} h_{0}$ and $h_{y}\equiv \partial_{y} h_{0}$ are indicated.}
\end{figure}

\begin{figure}[t!]
\begin{center}
\includegraphics[width=1.0\textwidth, angle=0.0, scale=0.6]{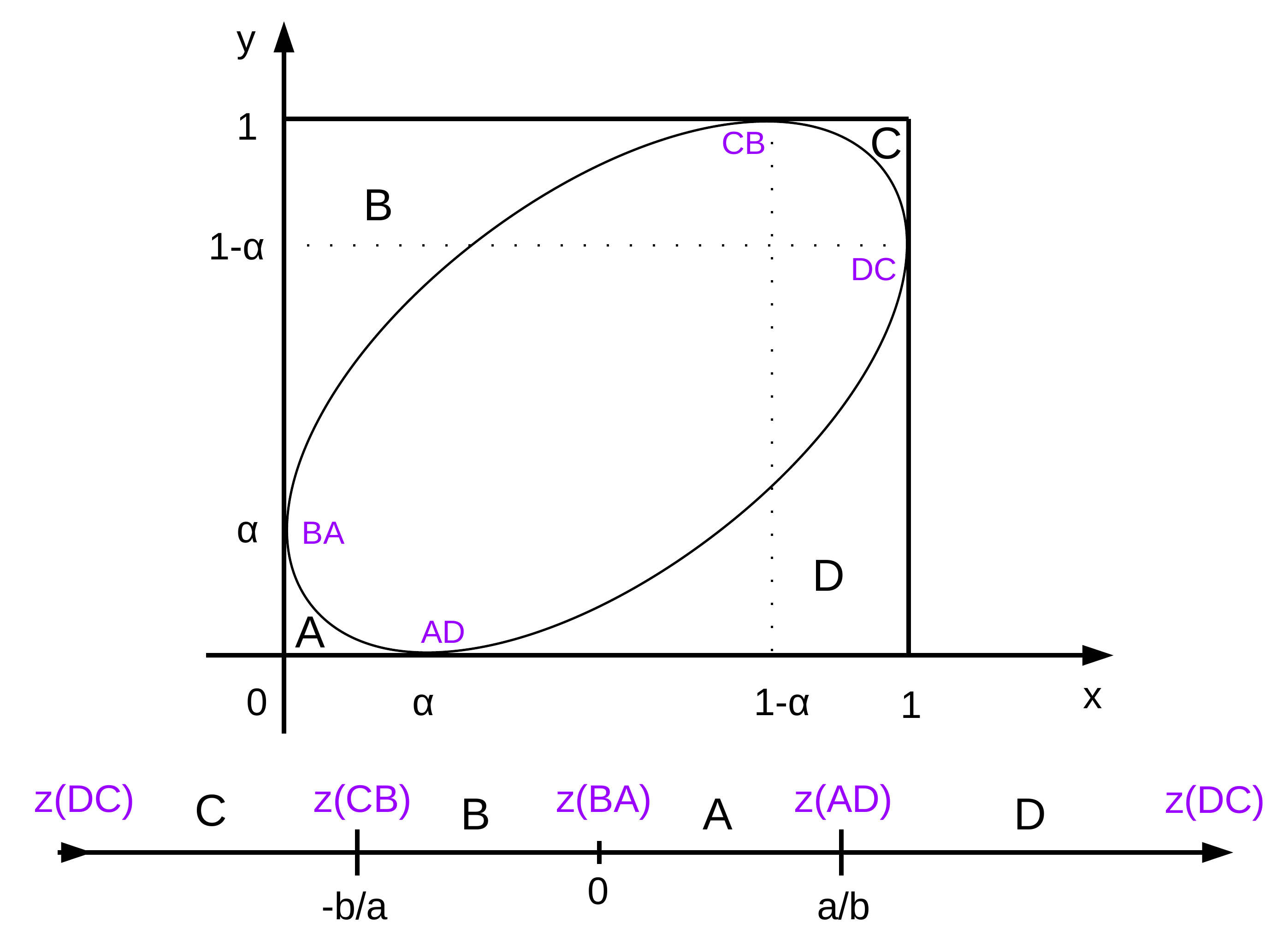}
\end{center}
\caption{\label{figAppendix2} The conformal mapping $z(x,y)$ maps the interior of the ellipse $\pa E$ to the upper half plane such that
points $DC$, $CB$, $BA$, $AD$ are mapped to points $\mp \infty, -b/a, 0, a/b$ respectively.}
\end{figure}

\subsection{}
Let us consider the asymptotical behavior of $|z|$ when $(x,y)\in E$ approaches $\pa E\cap D$.
We know that there $\pa_x h_{0} \to 0$, $\pa_yh_{0} \to 1$. Introduce variables $\delta_{x}=\pi h_{x}$ and $\delta_{y} = \pi - \pi h_{y}$.
At generic points of $\partial E$ the ratio $\delta_{x}/\delta_{y}$ is finite.
However, when $\partial E$ is tangent to the boundary of the domain (see points $DC$, $CB$, $BA$, $AD$ in Fig.~\ref{figAppendix1}) there is a change of the asymptotic of the height function at $\partial E$ and, as a consequence, $\delta_{x}/\delta_{y} \to 0$ or $\infty$.
Below, we analyze these asymptotics and check the signs in the formula for $|z|$.

In the current case, we have $\delta_{x},\delta_{y} \to +0$ as $(x,y)\to \partial E\cap D$, therefore
$$
\sin{(-\pi h_{y})}=\sin{(-\pi+\delta_{y})}\to-\delta_{y},
$$
$$
\sin{(\pi(h_{x}-h_{y}))}=\sin{(-\pi+\delta_{x}+\delta_{y})}\to-(\delta_{x}+\delta_{y}),
$$
$$
\sin{(\pi(h_{x}+h_{y}))}=\sin{(\pi+\delta_{x}-\delta_{y})}\to-(\delta_{x}-\delta_{y}).
$$
From here we derive the asymptotic of $|z|$ near the $D$-part of the boundary
$$
|z|=\frac{1}{2 \, a \, b \, \delta_{y}} \left( (a^{2}+b^{2}) \delta_{x} + (a^{2}-b^{2}) \delta_{y} \pm \sqrt{ (a^{2}+b^{2})^{2} (\delta_{x}^{2}+\delta_{y}^{2})+ 2 (a^{4}-b^{4}) \delta_{x} \delta_{y} } \right).
$$
We have an obvious inequality
$$
\left( (a^{2}+b^{2}) \delta_{x} + (a^{2}-b^{2}) \delta_{y} \right)^{2} < (a^{2}+b^{2})^{2} (\delta_{x}^{2}+\delta_{y}^{2})+ 2 (a^{4}-b^{4}) \delta_{x} \delta_{y}
$$
which means that one of the solutions is negative, the other is positive, corresponding to the plus sign in (\ref{zw}).
The positive solution is
$$
|z|=\frac{1}{2 \, a \, b \, \delta_{y}} \left( (a^{2}+b^{2}) \delta_{x} + (a^{2}-b^{2}) \delta_{y} + \sqrt{ (a^{2}+b^{2})^{2} (\delta_{x}^{2}+\delta_{y}^{2})+ 2 (a^{4}-b^{4}) \delta_{x} \delta_{y} } \right)=
$$
$$
\frac{a^2+b^2}{2ab} \left( \frac{\delta_{x}}{\delta_{y}} + \frac{a^2-b^2}{a^2+b^2} + \sqrt{1+ \left( \frac{\delta_{x}}{\delta_{y}} \right)^2 + 2 \frac{a^2-b^2}{a^2+b^2} \frac{\delta_{x}}{\delta_{y}} } \right).
$$
We have two boundary points of the segment $\pa E\cap D$, one is $AD=A\cap D=(0,\alpha)$, the other is 
$DC=C\cap D=(1,1-\alpha)$. As $(x,y)\to AD$ near $\pa E\cap D$, we have $\frac{\delta_{x}}{\delta_{y}} \to 0$, and therefore 
$$
|z| \to \frac{a^2+b^2}{2ab} \left(1 + \frac{a^2-b^2}{a^2+b^2} \right) = \frac{a}{b}=\frac{1}{\alpha}.
$$
When $(x,y)\to DC$ near $\pa E\cap D$, we have $\frac{\delta_{x}}{\delta_{y}} \to \infty$ and therefore $ |z| \to \infty$.
Thus, on $D \cap \partial E$, $z$ is real and 
$$
\frac{a}{b} < z < \infty.
$$
\subsection{}
Let us consider the asymptotical behavior of $|z|$ when $(x,y)\in E$ approaches $\pa E\cap C$.
As in the previous case, we define $\delta_x,\delta_y$ as $\delta_{x} = \pi - \pi h_{x}$, $\delta_{y}= \pi - \pi h_{y} $.
Then, since near the boundary $\delta_{x},\delta_{y} \to +0$,
$$
\sin{(-\pi h_{y})}=\sin{(-\pi+\delta_{y})}\to-\delta_{y},
$$
$$
\sin{(\pi(h_{x}-h_{y}))}=\sin{(\delta_{y}-\delta_{x})}\to\delta_{y}-\delta_{x},
$$
$$
\sin{(\pi(h_{x}+h_{y}))}\to-(\delta_{x}+\delta_{y}).
$$
For $|z|$, we obtain
$$
|z|=-\frac{1}{2 \, a \, b \, \delta_{y}} \left( a^{2} \left(\delta_{y}-\delta_{x} \right) - b^{2} \left(\delta_{x}+\delta_{y} \right) \mp \sqrt{ 4 a^2 b^2 \delta_{y}^{2} + \left( a^{2} \left(\delta_{x}-\delta_{y} \right) - b^{2} \left(\delta_{x}+\delta_{y} \right) \right)^{2} } \right)=
$$
$$
\frac{1}{2 \, a \, b \, \delta_{y}} \left( (a^{2}+b^{2}) \delta_{x} - (a^{2}-b^{2}) \delta_{y} \pm \sqrt{ (a^{2}+b^{2})^{2} (\delta_{x}^{2}+\delta_{y}^{2})- 4 (a^{4}-b^{4}) \delta_{x} \delta_{y} } \right).
$$
The inequality 
$$
\left( (a^{2}+b^{2}) \delta_{x} - (a^{2}-b^{2}) \delta_{y} \right)^{2} < (a^{2}+b^{2})^{2} (\delta_{x}^{2}+\delta_{y}^{2})- 4 (a^{4}-b^{4}) \delta_{x} \delta_{y}
$$
implies that the positive solution corresponds to the minus sign in (\ref{zw}) and its asymptotic is
$$
|z|=\frac{a^2+b^2}{2ab} \left( \frac{\delta_{x}}{\delta_{y}} - \frac{a^2-b^2}{a^2+b^2} + \sqrt{\left( \frac{\delta_{x}}{\delta_{y}} \right)^2 -4 (a^4-b^4) \frac{\delta_{x}}{\delta_{y}} + 1 } \right).
$$
When $(x,y)\to CB=(1-\alpha,1)$ near $\pa E\cap C$, we have $\frac{\delta_{x}}{\delta_{y}} \to 0$ and therefore 
$$
|z| \to \frac{a^2+b^2}{2ab} \left( 1 - \frac{a^2-b^2}{a^2+b^2} \right)=\frac{b}{a}.
$$
When $(x,y)\to DC=(1, 1-\alpha)$ near $\pa E\cap C$, we have $\frac{\delta_{x}}{\delta_{y}} \to \infty$ and therefore $|z| \to \infty $
Thus, on $C \cap \partial E$, $z$ is real and
$$
-\infty < z < -\frac{b}{a}.
$$
\subsection{}
Let us consider the asymptotical behavior of $|z|$ when $(x,y)\in E$ approaches $\pa E\cap B$.
Now define $\delta_x$ and $\delta_y$ as $\delta_{x}= \pi - \pi h_{x}$ and $\delta_{y}=\pi h_{y}$. Near the boundary 
$\delta_{x}, \delta_{y} \to +0$ and 
$\sin{(-\pi h_{y})}\to -\delta_{y}$, $\sin{(\pi(h_{x}-h_{y}))}=\sin{(\pi-\delta_{x}-\delta_{y})}\to \delta_{x}+\delta_{y}$,
$\sin{(\pi(h_{x}+h_{y}))}=\sin{(\pi-\delta_{x}+\delta_{y})}\to \delta_{x}-\delta_{y}$.
Thus, for
$z$
we obtain
$$
|z|=-\frac{1}{2 \, a \, b \, \delta_{y}} \left( a^{2} (\delta_{x}+\delta_{y}) + b^{2} (\delta_{x}-\delta_{y}) \mp \sqrt{ 4 a^{2} b^{2} \delta_{y}^{2}+ \left( a^{2} \left(\delta_{x}+\delta_{y} \right) + b^{2} \left(\delta_{x}-\delta_{y} \right) \right)^{2} } \right).
$$
The inequality
$$
(a^{2}+b^{2})^{2} \left(\delta_{x}^{2}+\delta_{y}^{2} \right) +2 (a^{4}-b^{4}) \delta_{x} \delta_{y}> \left( (a^{2}+b^{2}) \delta_{x} +(a^{2}-b^{2}) \delta_{y} \right)^2 
$$
implies that the positive solution corresponds to the plus sign, and for the asymptotic we have
$$
|z|=\frac{a^2+b^2}{2ab} \left( -\frac{\delta_{x}}{\delta_{y}} - \frac{a^2-b^2}{a^2+b^2} + \sqrt{ 1 + \left(\frac{\delta_{x}}{\delta_{y}} \right)^2 +2 \frac{a^2-b^2}{a^2+b^2} \frac{\delta_{x}}{\delta_{y}} } \right).
$$
When $(x,y)\to CB=(1-\alpha,1)$, we have $\frac{\delta_{x}}{\delta_{y}} \to 0$ and 
$$
|z| \to \frac{a^2+b^2}{2ab} \left( 1 - \frac{a^2-b^2}{a^2+b^2} \right)=\frac{b}{a}. 
$$
When $(x,y)\to BA=(0,\alpha)$ we have $\frac{\delta_{x}}{\delta_{y}} \to \infty$ and 
$$
|z|=O \left( \frac{\delta_y}{\delta_x} \right) \to 0.
$$
Thus, on $B \cap \partial E$ $z$ is real and
$$
-\frac{b}{a} < z < 0.
$$
\subsection{}
Let us consider the asymptotical behavior of $|z|$ when $(x,y)\in E$ approaches $\pa E\cap A$.
Define $\delta_{x}=\pi h_{x}$ and $\delta_{y}=\pi h_{y}$. Near the boundary $\delta_{x}, \delta_{y} \to +0$
and
$\sin{(-\pi h_{y})}\to -\delta_{y}$, $\sin{(\pi(h_{x}-h_{y}))}=\delta_{x}-\delta_{y}$,
$\sin{(\pi(h_{x}+h_{y}))}=\delta_{x}+\delta_{y}$.
For the asymptotic of $|z|$ in this region we obtain
$$
|z|=-\frac{1}{2 \, a \, b \, \delta_{y}} \left( a^{2} (\delta_{x}-\delta_{y}) + b^{2} (\delta_{x}+\delta_{y}) \mp \sqrt{ 4 a^{2} b^{2} \delta_{y}^{2}+ \left( a^{2} \left(\delta_{x}-\delta_{y} \right) + b^{2} \left(\delta_{x}+\delta_{y} \right) \right)^{2} } \right)=
$$
$$
-\frac{1}{2 \, a \, b \, \delta_{y}} \left( (a^{2}+b^{2}) \delta_{x} - (a^{2}-b^{2}) \delta_{y} \mp \sqrt{ (a^{2}+b^{2})^{2} (\delta_{x}^{2}+\delta_{y}^{2}) -4 (a^{4}-b^{4}) \delta_{x} \delta_{y} } \right).
$$
From the inequality 
$$
\left( (a^{2}+b^{2}) \delta_{x} -(a^{2}-b^{2}) \delta_{y} \right)^{2} < (a^{2}+b^{2})^{2} \left(\delta_{x}^{2}+\delta_{y}^{2}\right) -4 (a^{4}-b^{4}) \delta_{x} \delta_{y} 
$$
we conclude that the positive solution corresponds to the plus sign and
$$
|z|=\frac{1}{2ab} \left( a^2-b^2 - (a^2+b^2)\frac{\delta_{x}}{\delta_{y}} + \sqrt{ (a^2+b^2)^2 \left(1 + \left(\frac{\delta_{x}}{\delta_{y}} \right)^2 \right) -4 (a^{4}-b^{4}) \frac{\delta_{x}}{\delta_{y}}} \right).
$$
When $(x,y)\to BA=(0, \alpha)$ near $\pa E\cap A$ we have $\frac{\delta_{x}}{\delta_{y}} \to 0$ and, therefore, $ |z| \to \frac{a}{b}$
When $(x,y)\to AD=(\alpha,0)$ near $\pa E\cap A$ we have 
$\frac{\delta_{x}}{\delta_{y}} \to \infty$ and $|z| \to 0$.
Thus, on $A \cap \partial E$, $z$ is real and
$$
0 < z < \frac{a}{b}.
$$

The obtained mapping is depicted in Fig.~\ref{figAppendix2}.


\end{document}